\documentclass[12pt]{elsarticle}

\usepackage[american]{babel}
\usepackage[utf8]{inputenc}
\usepackage{ae,aecompl}
\usepackage{bm}
\usepackage{url}
\usepackage{color}
\usepackage{booktabs}
\usepackage{subfigure}

\usepackage{amsmath} 
\usepackage{amssymb}
\usepackage{commath}
\usepackage{dsfont}
\usepackage[bbgreekl]{mathbbol}

\everymath={\displaystyle}

\newcommand{\R}{\ensuremath{\mathbb{R}}}           
\newcommand{\N}{\ensuremath{\mathbb{N}}}           

\renewcommand{\vec}[1]{\mathbf{#1}}

\newcommand{\mat}[1]{\mathbf{#1}}

\newcommand{\transp}[1]{#1^{\scriptscriptstyle T}}

\renewcommand{\SS}{\Theta}

\newcommand{\SSpt}{\theta}

\newcommand{\SF}{\mathbb{\Sigma}}

\newcommand{\PM}{\mathbb{P}}

\newcommand{\randvar}[1]{\mathbb{#1}}

\newcommand{\cdf}[2]{F_{\tiny{#1}}(#2)}

\newcommand{\expval}[1]{E \left\lbrace   #1 \right\rbrace  }

\newcommand{\mean}[1]{\mu_{#1}}

\newcommand{\var}[1]{\sigma^2_{#1}}

\newcommand{\stddev}[1]{\sigma_{#1}}

\newcommand{\autocov}[1]{C_{#1}}

\begin{document}

\begin{frontmatter}

\journal{Journal of Sound and Vibration}

\title{Quantification of parametric uncertainties\\ induced by irregular soil loading in orchard\\ tower sprayer nonlinear dynamics}

\author[uerj]{Americo Cunha Jr\corref{cor}}
\ead{americo@ime.uerj.br}

\author[uffs]{Jorge Luis Palacios Felix}
\ead{jorge.felix@uffs.edu.br}

\author[unesp,ita]{Jos\'{e} Manoel Balthazar}
\ead{jmbaltha@ita.br}

\cortext[cor]{Corresponding author.}

\address[uerj]{Universidade do Estado do Rio de Janeiro}

\address[uffs]{Universidade Federal da Fronteira Sul}

\address[unesp]{UNESP-  S\~{a}o Paulo State University}

\address[ita]{Instituto Tecnol\'{o}gico de Aeron\'{a}utica}

\date{}

\begin{abstract}

This paper deals with the nonlinear stochastic dynamics of an orchard 
tower sprayer subjected to random excitations due to soil irregularities. 
A consistent stochastic model of uncertainties is constructed to describe 
random loadings and to predict variabilities in mechanical system response.
The dynamics is addressed in time and frequency domains. Monte Carlo 
method is employed to compute the propagation of uncertainties 
through the stochastic model. Numerical simulations reveals a very rich 
dynamics, which is able to produce chaos. This numerical study also indicates 
that lateral vibrations follow a direct energy cascade law. A probabilistic 
analysis reveals the possibility of large lateral vibrations during the 
equipment operation.
\end{abstract}

\begin{keyword} 
orchard tower sprayer, nonlinear dynamics, uncertainty quantification,
parametric probabilistic approach, Karhunen-Lo\`{e}ve decomposition
\end{keyword}

\end{frontmatter}

\section{Introduction}

The proliferation of pests in agricultural industry can be harmful to 
consumers and producers, since it can cause problems such as a
reduction in the products quality, partial/total loss of the plantation, etc. 
Thus, the process of agricultural spraying for pest control
is of great importance in orchards, vegetable gardens, etc.
In general, the spraying of orchards is done with the aid of an
equipment called \emph{tower sprayer}, that consists of a 
reservoir and several fans mounted on an articulated tower, 
which is supported by a vehicle suspension \cite{sartoriJunior2009p417}. 
Due to soil irregularities this equipment is subjected to loads
of random nature, which may hamper the fluid spraying proper dispersion.

Primary studies on this topic are presented in
\cite{sartoriJunior2009p417,sartorijunior2008,sartorijunior2007}, 
using a mathematical model that considers an inverted pendulum mounted on a moving base 
to emulate the equipment. These works perform deterministic analyzes to investigate 
the influence of certain parameters (stiffness, torsional damping, etc)
in the model response. In addition, references \cite{sartorijunior2008} and
\cite{sartorijunior2007} present a detailed study of the associated linear dynamics.
In all cases, the observed behavior is physically reasonable, but also the analyzes 
are limited to simple situations, once the model 
does not take into account the system dynamics underlying uncertainties. 
In fact, system parameters have uncertainties 
due to a series of factors such as variabilities intrinsic to the 
manufacturing process, materials and geometric imperfections, etc
\cite{soize2012,soize2013p2379}. Taking such uncertainties into 
account is essential for making robust predictions, but also, it has been
becoming a common practice in engineering
\cite{cunhajr2015p849,cunhajr2015p809,perrin2015p945,ritto2015p101,beck2015p479}.

In this sense, this paper aims to construct a consistent stochastic model
to describe the nonlinear dynamics of an orchard tower 
sprayer, taking parametric uncertainties into account. In a first analysis, 
the authors concentrate their efforts in tires excitation uncertainties, 
induced by soil irregularities, once these loads are extremely 
complex and have great influence in the system dynamics. For this
purpose, it is more realistic to describe the system dynamics by means of
a probabilistic model of uncertainties, since in this type of approach 
uncertainties are naturally characterized \cite{soize2012}.
Some initiatives in this direction were presented by the authors in 
two conference papers \cite{cunhajr2015cobem,cunhajr2016uncertainties}, 
where a harmonic random process was used to emulate the aleatory loadings. 
But now, they intend to construct the random excitations using 
Karhunen-Lo\`{e}ve (KL) decomposition, seeking a better characterization 
of the loads. This work also intends to deeply investigate in depth the 
effects of random excitation in the tower sprayer response, 
and compute the probability of undesirable operating events, such as 
large lateral vibrations.

The rest of this paper is organized as follows. In section~2,
it is presented a deterministic model to describe the sprayer 
nonlinear dynamics. A stochastic model to to take into account 
the uncertainties associated with the soil induced loading 
is shown in section~3. The results of the numerical experiments 
conducted in this work are presented and discussed in section~4. 
Finally, in section~5, the main conclusions are highlighted,
and some paths for future works are indicated.

\section{Deterministic modeling}
\label{determ_model}

\subsection{Physical system definition}
\label{def_phys_sys}

\begin{figure}[h]
	\centering
	\includegraphics[scale=0.7]{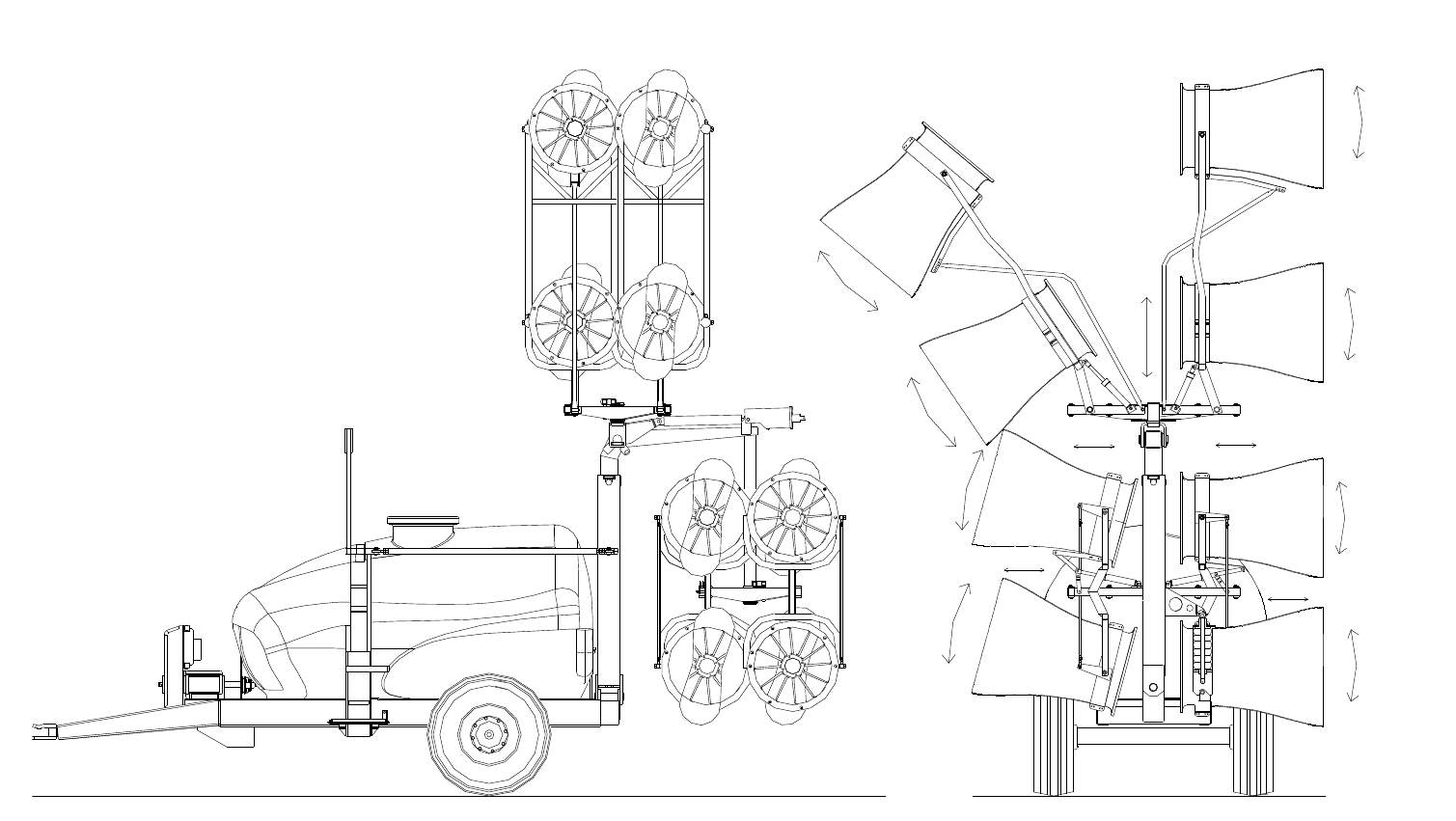}
	\caption{Schematic representation of the tower sprayer.
	Adapted from \cite{sartorijunior2008} and courtesy of M\'{a}quinas Agr\'{\i}colas Jacto S/A.} 
	\label{mech_sys_fig}
\end{figure}

The mechanical system of interest here is the tower sprayer
schematically represented in Figure~\ref{mech_sys_fig}.
It consists of a reservoir tank, used to store a spraying
fluid, which is mounted onto a vehicular suspension. 
In this suspension, there is a support tower where sixteen 
fans are arranged in columns, eight on the right and eight 
pointing to left. These fans are used to pulverize 
an orchard. As this equipment moves through a rough terrain, 
vertical and horizontal vibrations may be observed.

\subsection{Physical system parameterization}
\label{def_phys_sys}

\begin{figure}[h!]
	\centering
	\subfigure[]{\includegraphics[scale=0.3]{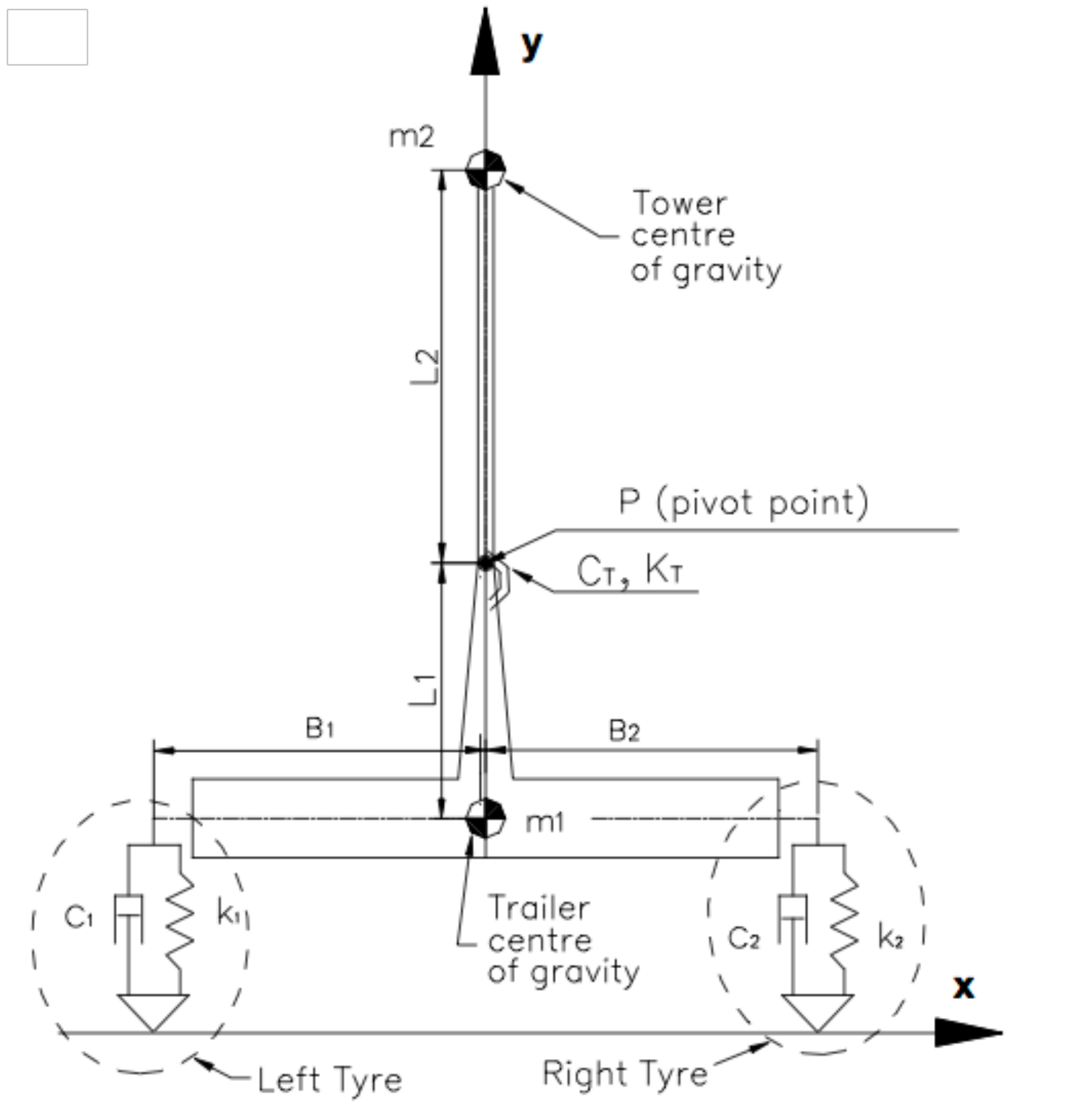}} ~~
	\subfigure[]{\includegraphics[scale=0.3]{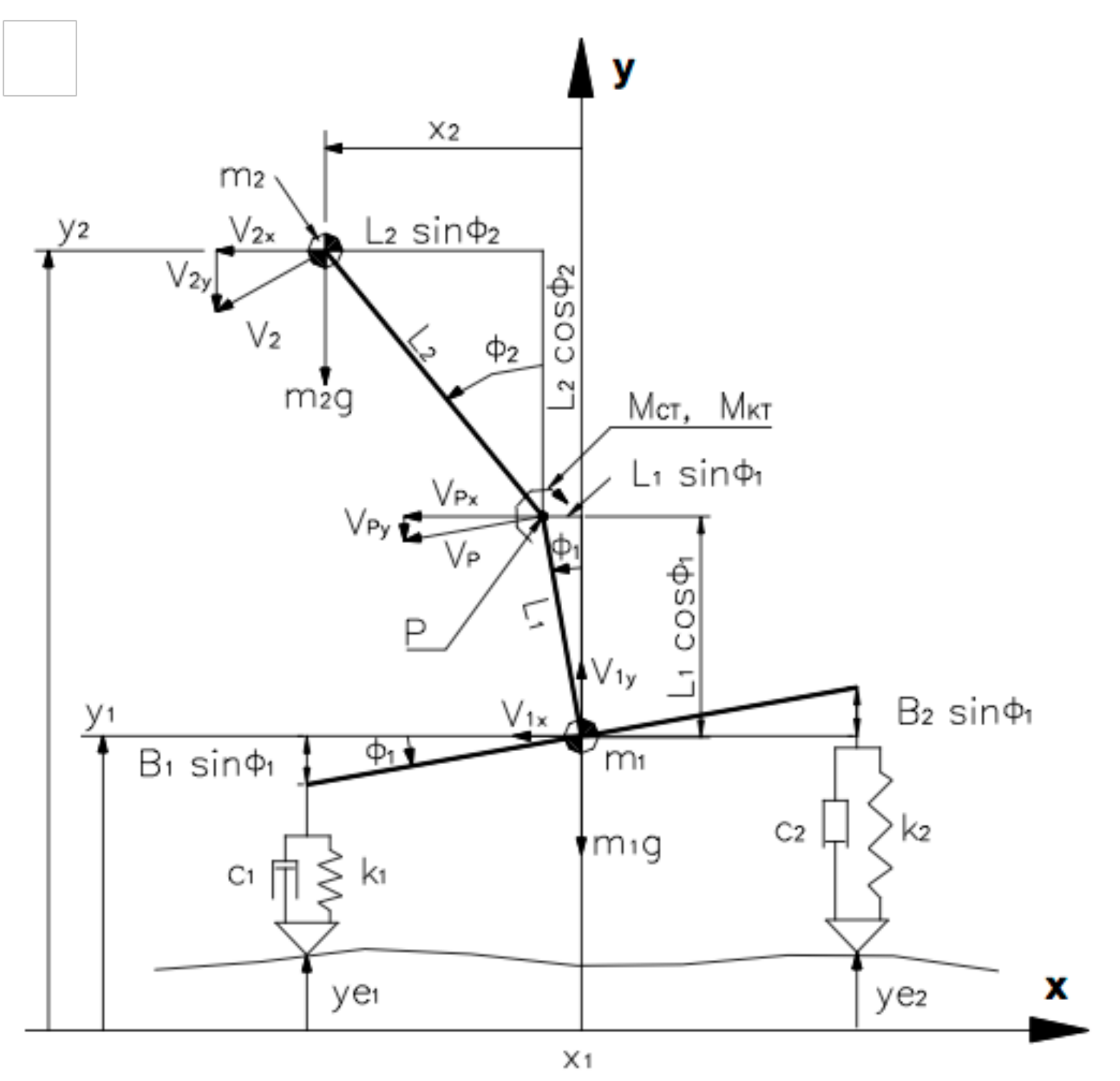}}
	\caption{Schematic representation of the mechanical-mathematical model 
	for the tower sprayer: an inverted pendulum mounted on a rigid trailer.
	(a) Static equilibrium configuration; (b) Off equilibrium configuration.
	Adapted from \cite{sartorijunior2008}.}
	\label{double_pendulum_model}
\end{figure}

For modeling purposes the orchard sprayer tower is
considered as the multibody system illustrated in 
Figure~\ref{double_pendulum_model}, such as proposed by 
\cite{sartoriJunior2009p417,sartorijunior2008}. Suspension 
chassis and reservoir tank are emulated by a rigid trailer 
with mass $m_1$. The vertical tower and funs are 
modeled by an inverted rigid pendulum of mass $m_2$.
Their moments of inertia, with respect to their center of gravity, 
are respectively denoted by $I_1$ and $I_2$. The point of 
articulation between the trailer and tower, denoted by $P$, has 
torsional stiffness $k_T$ and damping torsional coefficient $c_T$. 
Its distance to the trailer center of gravity is $L_1$ and the pendulum 
arm length is dubbed $L_2$. The left wheel of the vehicle suspension,
located at a distance $B_1$ from trailer center line, is represented by 
a pair spring/damper with constants $k_1$ and $c_1$, respectively, 
and it is subject to a vertical displacement $y_{e1}$. Similarly, the right wheel 
is represented by a pair spring/damper characterized by $k_2$ and $c_2$, 
it is $B_2$ away from the trailer center line, and it displaces vertically $y_{e2}$.
For simplicity, the sprayer translational velocity is supposed to be a
constant v. The acceleration of gravity is denoted by $g$.

Introducing the inertial frame of reference $XY$, the horizontal and vertical 
displacements of the trailer center of mass are respectively measured by 
$x_1$ and $y_1$, while its rotation is computed by $\phi_1$. The horizontal 
and vertical displacements of the tower center of mass are given by $x_2$ and 
$y_2$, respectively, and its rotation with respect to the trailer is denoted by $\phi_2$. 

As the trailer horizontal movement is limited by the pair of tires, one has $x_1 = 0$.
It can also be deduced from the geometry of Figure~\ref{double_pendulum_model} 
that

\begin{equation}
	x_2 = -L_1 \, \sin \phi_1 - L_2 \, \sin \phi_2,
\end{equation}

\noindent
and

\begin{equation}
	y_2 = y_1 + L_1 \, \cos \phi_1 + L_2 \, \cos \phi_2.
\end{equation}

\noindent
Therefore, this model has degrees of freedom: $y_1$, $\phi_1$ and $\phi_2$.
The main \emph{quantity of interest} (QoI) in the study of this dynamics
is the time function $x_2$.

\subsection{Lagrangian formalism}
\label{lagrange_form}

Euler-Lagrange equations are employed to obtain the system dynamics

\begin{equation}
  \dpd{}{t} \left(\dpd{\mathcal{T}}{\dot{q}} \right) - \dpd{\mathcal{T}}{q}  
  + \dpd{\mathcal{V}}{q}  + \dpd{\mathcal{D}}{\dot{q}} = 0,~~~q=\{y_1,\phi_1,\phi_2\},
\end{equation}

\noindent
where the upper dot is an abbreviation for time derivative, and the functionals
of kinetic energy, potential energy and dissipation are, respectively, given by
	
\begin{equation}
	\mathcal{T} = \frac{1}{2} m_1 \, \dot{y}_1 +
							\frac{1}{2} m_2 \left( \dot{x}_2 + \dot{y}_2 \right) +
	\frac{1}{2} I_1 \dot{\phi}_1 + \frac{1}{2} I_2 \dot{\phi}_2,
\end{equation}
	
\begin{equation}
\begin{split}
	\mathcal{V} = m_1 \, g\, y_1 + m_2 \, g\, y_2 +
							\frac{1}{2} k_1 \left( y_1 - B_1\,\sin{\phi_1} - y_{e1} \right)^2 +\\
							\frac{1}{2} k_2 \left( y_1 + B_2\,\sin{\phi_1} - y_{e2} \right)^2 +
							\frac{1}{2} k_T \left( \phi_2 - \phi_1 \right)^2,
\end{split}
\end{equation}

\noindent
and

\begin{equation}
\begin{split}
	\mathcal{D} = \frac{1}{2} c_1 \left( \dot{y}_1 - B_1\,\dot{\phi}_1\,\cos{\phi_1} - y_{e1} \right)^2 +\\
							\frac{1}{2} c_2 \left( \dot{y}_1 + B_2\,\dot{\phi}_1\,\cos{\phi_1} - y_{e1} \right)^2 +
							\frac{1}{2} c_T \left( \dot{\phi}_2 - \dot{\phi}_1 \right)^2.
\end{split}
\end{equation}

After the calculation, the following set of nonlinear ordinary differential
equations is obtained

\begin{eqnarray}
\resizebox{0.92\hsize}{!}{$
			\mat{M} \left( \begin{array}{c} 	\ddot{y}_1(t)\\ \ddot{\phi}_1(t)\\	\ddot{\phi}_2(t)\\\end{array} \right) +
			\mat{N} \left( \begin{array}{c} 	\dot{y}^2_1(t)\\ \dot{\phi}^2_1(t)\\	\dot{\phi}^2_2(t)\\\end{array} \right) +
			\mat{C} \left( \begin{array}{c} 	\dot{y}_1(t)\\ \dot{\phi}_1(t)\\	\dot{\phi}_2(t)\\\end{array} \right) +
			\mat{K} \left( \begin{array}{c} 	y_1(t)\\ \phi_1(t)\\	\phi_2(t)\\\end{array} \right) = \vec{g} - \vec{h},
			$}
			\label{eq_motion}
\end{eqnarray}

\noindent
where $\mat{M}$, $\mat{N}$, $\mat{C}$ and $\mat{N}$ are
$3 \times 3$ (configuration dependent) real matrices, 
respectively, defined by

\begin{eqnarray}
\resizebox{0.92\hsize}{!}{$
		\mat{M} =
		\left[ \begin{array}{ccc}
		m_1+m_2                           & -m_2 \, L_1 \, \sin \phi_1                                        & -m_2 \, L_2 \, \sin \phi_1 \\
		-m_2 \, L_1 \, \sin \phi_1 & I_1+m_2 \, L_1^2                                                      & m_2 \, L_1 \, L_2 \, \cos \left(\phi_2-\phi_1\right) \\
		-m_2 \, L_2 \, \sin \phi_1 & m_2 \, L_1 \, L_2 \, \cos \left(\phi_2-\phi_1\right) & I_2+m_2 \, L_2^2 \\
		\end{array} \right],
		$}
		\label{matrix_M_eq}
\end{eqnarray}

\begin{eqnarray}
		\mat{N} =
		\left[ \begin{array}{ccc}
		0 & -m_2 \, L_1 \, \cos \phi_1   & -m_2 \, L_2 \, \cos \phi_2 \\
		0 & 0                                                     & - m_2 \, L_1 \, L_2 \, \sin \left(\phi_2-\phi_1\right) \\
		0 & - m_2 \, L_1 \, L_2 \, \sin \left(\phi_2-\phi_1\right) & 0 \\
		\end{array} \right],
		\label{matrix_N_eq}
\end{eqnarray}

\begin{eqnarray}
		\mat{C} =
		\left[ \begin{array}{ccc}
		c_1+c_2                                     & (c_2\,B_2-c_1\,B_1) \cos \phi_1                        & 0 \\
		(c_2\,B_2-c_1\,B1) \cos \phi_1 & c_T + (c_1\,B^2_1+c_2\,B^2_2) \cos^2 \phi_1  & - c_T \\
		0                                                & - c_T                                                                    & c_T \\
		\end{array} \right],
		\label{matrix_C_eq}
\end{eqnarray}

\noindent
and

\begin{eqnarray}
		\mat{K} =
		\left[ \begin{array}{ccc}
		k_1+k_2                                     & 0         & 0 \\
		(k_2\,B_2-k_1\,B1) \cos \phi_1 & k_T     & - k_T \\
		0                                                 & - k_T  & k_T \\
		\end{array} \right],
		\label{matrix_K_eq}
\end{eqnarray}

\noindent
and $\vec{g}$ and $\vec{h}$ are (configuration dependent) vectors in $\R^{3}$,
respectively, defined by

\begin{eqnarray}
		\vec{g} =
		\left( \begin{array}{c}
			 (k_2 \, B_2-k_1 \, B_1) \sin \phi_1 +  (m_1+m_2) g\\
			 	(k_1\,B_1^2+k_2\,B_2^2)\sin \phi_1 \, \cos \phi_1 - m_2 \, g \, L_1 \, \sin \phi_1\\
			 - m_2 \, g \, L_2 \, \sin \phi_2\\
		\end{array} \right),
\end{eqnarray}

\noindent
and

\begin{eqnarray}
\resizebox{0.92\hsize}{!}{$
		\vec{h} =
		\left( \begin{array}{c}
			 k_1 \, y_{e1} + k_2 \, y_{e2} + c_1 \, \dot{y}_{e1} + c_2  \, \dot{y}_{e2}\\
			 -k_1 \, B_1 \, \cos \phi_1 \, y_{e1} + k_2 \, B_2 \, \cos \phi_1 \, y_{e2}	
			 -c_1 \, B_1 \, \cos \phi_1 \, \dot{y}_{e1}	+c_2 \, B_2 \, \cos \phi_1 \, \dot{y}_{e2}\\
			 0\\
		\end{array} \right).
		$}
\end{eqnarray}


The deduction of Eq.(\ref{eq_motion}) can be seen in detail in \cite{sartorijunior2008}, 
which also obtain this set of equations through a Newtonian formulation.

\subsection{Static equilibrium configuration}

The static equilibrium configuration for the tower sprayer,
illustrated in Figure~\ref{double_pendulum_model}(a), 
and defined by

\begin{equation}
    y_1(0) = -\frac{(m_1+m_2)}{k_1+k_2}\,g,
    \qquad
    \phi_1(0) =  0,
    \qquad
    \phi_2(0) =  0,
	\label{ic_eq1}
\end{equation}

\noindent
and

\begin{equation}
    \dot{y}_1(0) =  0,
    \qquad
   \dot{\phi}_1(0) =  0,
    \qquad
    \dot{\phi}_2(0) =  0.
	\label{ic_eq2}
\end{equation}

\noindent
is assumed as the initial state of the system.
This is a stable equilibrium, where the system presents neither velocity
nor any rotation, but has a negative vertical displacement with 
respect to the level of reference.

\subsection{Nonlinear initial value problem}

By means of the generalized displacement $\vec{q}: t \in \R \mapsto \vec{q}(t) \in \R^6$,
the initial displacement vector $\vec{q}_0 \in \R^6$, and the nonlinear mapping 
$\vec{f}: \left( t,\vec{q}(t) \right) \in \R \times \R^6 \mapsto \vec{f} \left( \vec{q}(t) \right) \in \R^6$,
where

\begin{eqnarray}
		\vec{q}(t) =
		\left( \begin{array}{c}
			 y_1(t)\\
			 \phi_1(t)\\
			 \phi_2(t)\\
			 \dot{y}_1(t)\\
			 \dot{\phi}_1(t)\\
			 \dot{\phi}_2(t)\\
		\end{array} \right),
		\qquad
		\vec{q}_0 =
		\left( \begin{array}{c}
			 y_1(0)\\
			 \phi_1(0)\\
			 \phi_2(0)\\
			 \dot{y}_1(0)\\
			 \dot{\phi}_1(0)\\
			 \dot{\phi}_2(0)\\
		\end{array} \right),
\end{eqnarray}

\noindent
and

\begin{eqnarray}
\resizebox{0.92\hsize}{!}{$
		\vec{f} \left( \vec{q}(t)\right) =
		\left( \begin{array}{c}
			 \dot{y}_1(t)\\
			 \dot{\phi}_1(t)\\
			 \dot{\phi}_2(t)\\
			 - \mat{M}^{-1} \left( \mat{N} \left( \begin{array}{c} 	\dot{y}^2_1(t)\\ \dot{\phi}^2_1(t)\\	\dot{\phi}^2_2(t)\\\end{array} \right) +
			 								   \mat{C} \left( \begin{array}{c} 	\dot{y}_1(t)\\ \dot{\phi}_1(t)\\	\dot{\phi}_2(t)\\\end{array} \right) +
			 									\mat{K} \left( \begin{array}{c} 	y_1(t)\\ \phi_1(t)\\	\phi_2(t)\\\end{array} \right) - \vec{g} + \vec{h}
			 						\right)
		\end{array} \right),
  $ }
\end{eqnarray}

\noindent
it is possible write the dynamical system of
Eqs.(\ref{eq_motion}),  (\ref{ic_eq1}) and (\ref{ic_eq2})  as

\begin{equation}
		\vec{\dot{q}}(t)	= \vec{f} \left( \vec{q}(t) \right),
		\qquad
		\vec{q}(0) =  \vec{q}_{0},
		\label{ivp_eq}
\end{equation}

\noindent
a nonlinear initial value problem that is integrated using 
Runge-Kutta-Fehlberg method (RKF45) 
\cite{fehlberg1969,ascher2011}.



\section{Stochastic modeling}
\label{math_model}

\subsection{Aleatory nature of a tire displacement}

\begin{figure}[h]
	\centering
	\subfigure[left tire path]{\includegraphics[scale=0.82]{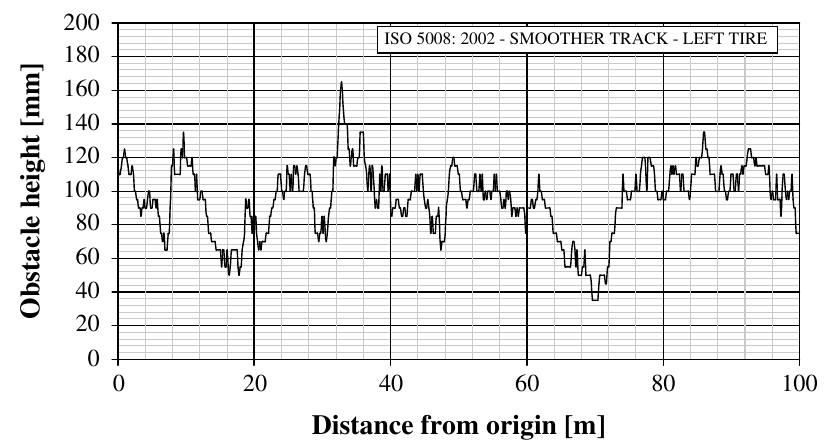}}
	\subfigure[right tire path]{	\includegraphics[scale=0.75]{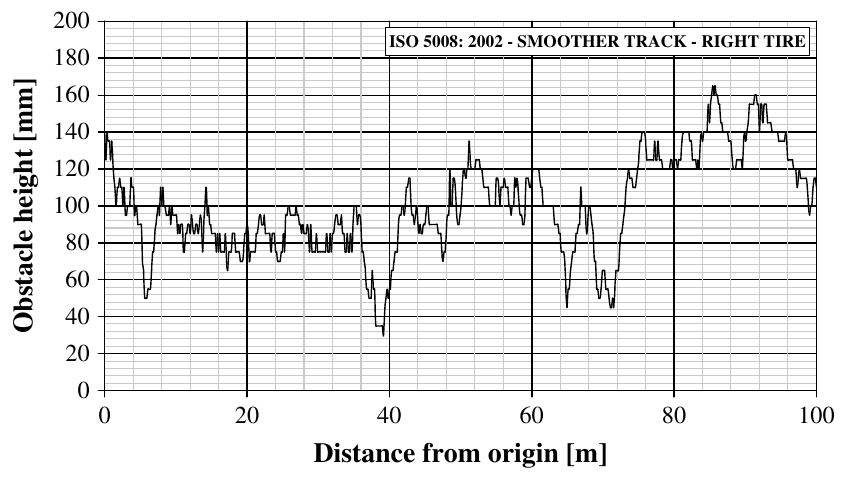}}
	\caption{Illustration of typical paths followed by sprayer 
	tower tires during operation. Adapted from \cite{sartoriJunior2009p417}.}
	\label{tires_path_fig}
\end{figure}

Typical paths followed by sprayer tower during its 
operation are illustrated in Figure~\ref{tires_path_fig}, which shows
tires vertical displacement as function of the traveled distance.
Note that sprayer tires undergo irregular displacements, which 
resembles a random signal not a smooth function. In this way, it is
better to describe the irregular form of tires displacement, and
therefore, the sprayer tower dynamics, with a stochastic dynamic model.

\subsection{Probabilistic framework}

In this work the mechanical system stochastic dynamics is
described through a parametric probabilistic approach 
\cite{soize2012,soize2013p2379},
which uses the probability space $(\SS, \SF, \PM)$, being $\SS$ 
the sample space, $\SF$ a $\sigma$-field over $\SS$, and 
$\PM: \SF \to [0,1] $ the probability measure.
Within this framework, the mathematical expectation operator is defined by

\begin{equation}
	\expval{\randvar{y}} = \int_{\R} y \, d \cdf{\randvar{y}}{y} \, ,
	\label{def_expval_op}
\end{equation}

\noindent
where $\randvar{y}$ is a real-valued random variable defined in
$(\SS, \SF, \PM)$, with probability distribution $\cdf{\randvar{y}}{y}$. 
With the aid of Eq.(\ref{def_expval_op}) it is possible to define statistics 
of $\randvar{y}$ such as mean value 
$\mean{\randvar{y}} = \expval{\randvar{y}}$, variance 
$\var{\randvar{y}} = \expval{\left( \randvar{y} - \mean{\randvar{y}} \right)^2}$,
and standard deviation $\stddev{\randvar{y}} = \sqrt{\var{\randvar{y}}}$.
Note that for random processes, which are ``time-dependent random variables",
these statistics present time dependence. Furthermore, the covariance function
of random process $\left\lbrace \randvar{y}(t), t \in \R \right\rbrace$, 
at time instants $t_1$ and $t_2$,
is defined by $\autocov{\randvar{y}}(t_1,t_2) = \expval{\left( \randvar{y}(t_1) - \mean{\randvar{y}}(t_1) \right) \left( \randvar{y}(t_2) - \mean{\randvar{y}}(t_2) \right)}$.


\subsection{Tire displacement modeling}
\label{tires_model}

The tire displacements have aleatory nature and present time dependence, 
so that they can be described by square-integrable random processes
$\left\lbrace \randvar{y}_{e1}(t), t \in \R \right\rbrace$ and
$\left\lbrace \randvar{y}_{e2}(t), t \in \R \right\rbrace$.
Accordingly, the trajectories illustrated in Figure~ \ref{tires_path_fig} 
can be thought as sample paths associated to these processes.

The dynamic behavior of one tire certainly influences
the way other tire behaves, i.e., there is some dependence between 
the two random processes. However, for lack of better knowledge
about the correlation between $\randvar{y}_{e1}(t)$ and 
$\randvar{y}_{e2}(t)$, these random processes are assumed 
to be independent. For convenience, they are also assumed to be stationary,
which implies that the means values $\mean{\randvar{y}_{e1}}$ 
and $\mean{\randvar{y}_{e2}}$ are constant, as well as
the standard deviations $\stddev{\randvar{y}_{e1}}$ and 
$\stddev{\randvar{y}_{e2}}$.

Once the tire displacement at certain instant of time 
has little influence on the value of this kinematic parameter at a distant time,
it is also reasonable to assume that covariance functions
of these processes present exponentially decaying behavior, i.e.,

\begin{equation}
	\autocov{\randvar{y}_{e1}}(t_1,t_2) = 
	\autocov{\randvar{y}_{e2}}(t_1,t_2) = 
	\exp{\left( - \frac{t_2-t_1}{a_{corr}/v} \right)} \, ,
	\label{autocov_ye1}
\end{equation}

\noindent
where $v$ is the translational velocity of the sprayer tower 
(supposed as constant) and $a_{corr}$ is a correlation length
for the processes $\randvar{y}_{e1}(t)$ and $\randvar{y}_{e2}(t)$.


\subsection{Random processes representation}

From the theoretical point of view, random processes
$\randvar{y}_{e1}(t)$ and $\randvar{y}_{e2}(t)$
are well defined with the information given
in section~\ref{tires_model}. However, 
for computational implementation purposes, it is necessary 
to represent these random processes (infinite-dimensional objects) 
in terms of a finite number of random variables \cite{xiu2010}.

This task can be efficiently done through the truncation of 
Karhunen-Lo\`{e}ve (KL) decomposition \cite{ghanem2003,xiu2010}, 
which is a powerful tool to represent random fields/processes
\cite{bellizzi2006p774,sampaio2007p22,stefanou20072p465,
bellizzi2009p491,bellizzi2009p1218,bellizzi2012p3509,bellizzi2015p245}.

KL expansion of $\left\lbrace \randvar{y}(t), t \in \R \right\rbrace$ writes as

\begin{equation}
		\randvar{y}(t) = \mean{\randvar{y}}(t) +
		\sum_{n=1}^{+\infty} \sqrt{\lambda_n} \, \varphi_n(t) \, \randvar{Y}_n,
		\label{kl_decomp}
\end{equation}

\noindent
where the pairs $\left( \lambda_n, \varphi_n \right)$ are solution of 
Fredholm integral equation

\begin{equation}
		\int_{\R} \autocov{{y}(t)} (t,s) \, \varphi_n (s) \, ds =
		\lambda_n \, \varphi_n (t), \qquad t \in \R,
\end{equation}

\noindent
and $\left\lbrace \randvar{Y}_n \right\rbrace_{n=1}^{+\infty}$ is
a family of zero-mean mutually uncorrelated random variables, i.e.,

\begin{equation}
		\mean{\randvar{Y}_n} = 0,
		\qquad \mbox{and} \qquad
		\expval{\randvar{Y}_n \randvar{Y}_m} = \delta_{mn}.
		\label{kl_uncorrelated}
\end{equation}

The approximation is obtained after the truncation of Eq.(\ref{kl_decomp}), i.e.,

\begin{equation}
		\randvar{y}(t) \approx \mean{\randvar{y}}(t) +
		\sum_{n=1}^{N_{KL}} \sqrt{\lambda_n} \, \varphi_n(t) \, \randvar{Y}_n,
		\label{kl_truncated}
\end{equation}

\noindent
where the integer $N_{KL}$ is chosen such that

\begin{equation}
		\frac{\sum_{n=1}^{N_{KL}} \lambda_n}{\sum_{n=1}^{+ \infty} \lambda_n} \geq \tau,
		\label{kl_truncated}
\end{equation}

\noindent
with $\tau = 99.9\%$, such as suggested by \cite{trindade2005p1015}.

\begin{figure}[h]
	\centering
	\subfigure[$\randvar{y}_{e1}$ path (left tire)]{\includegraphics[scale=0.35]{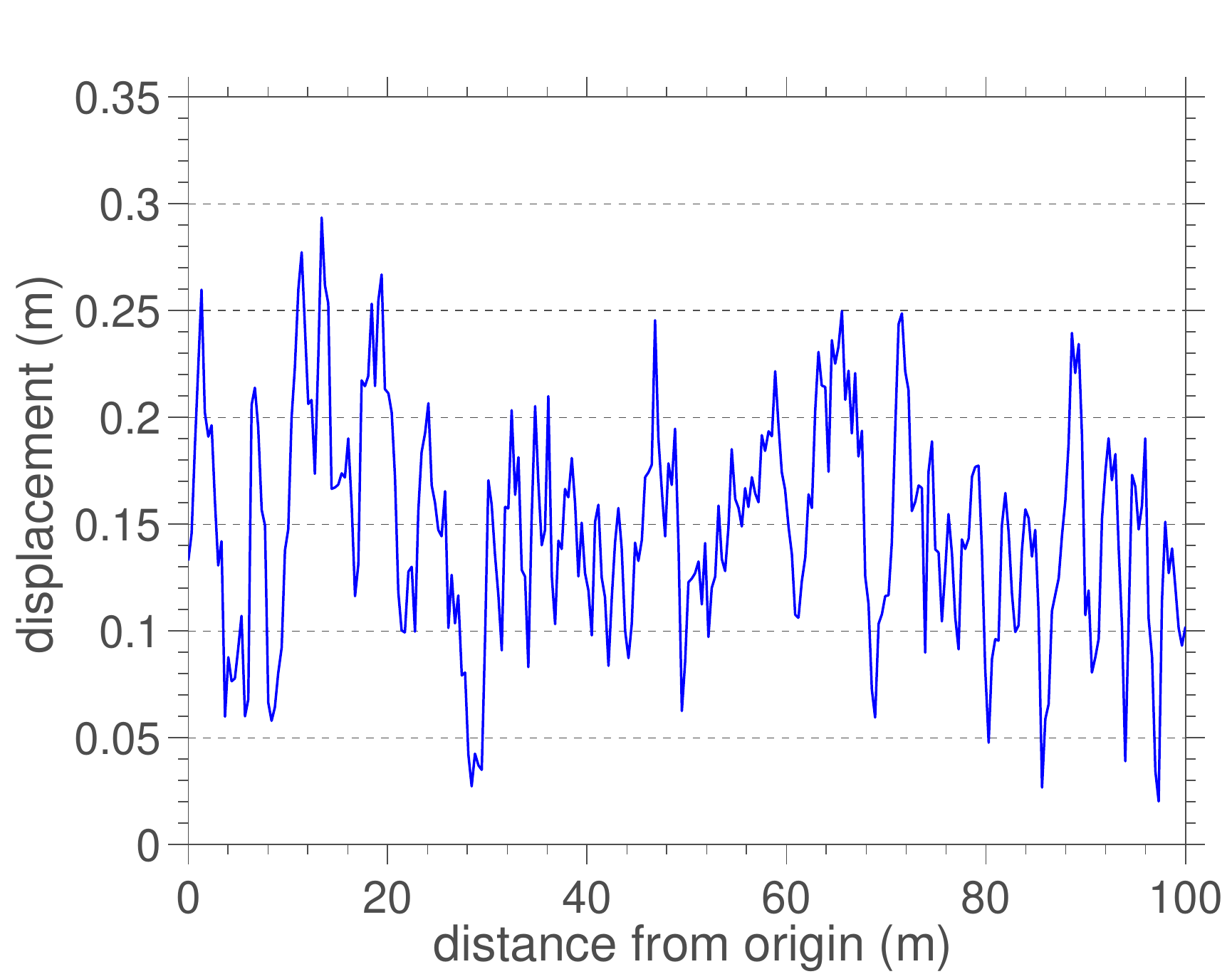}}
	\subfigure[$\randvar{y}_{e2}$ path (right tire)]{\includegraphics[scale=0.35]{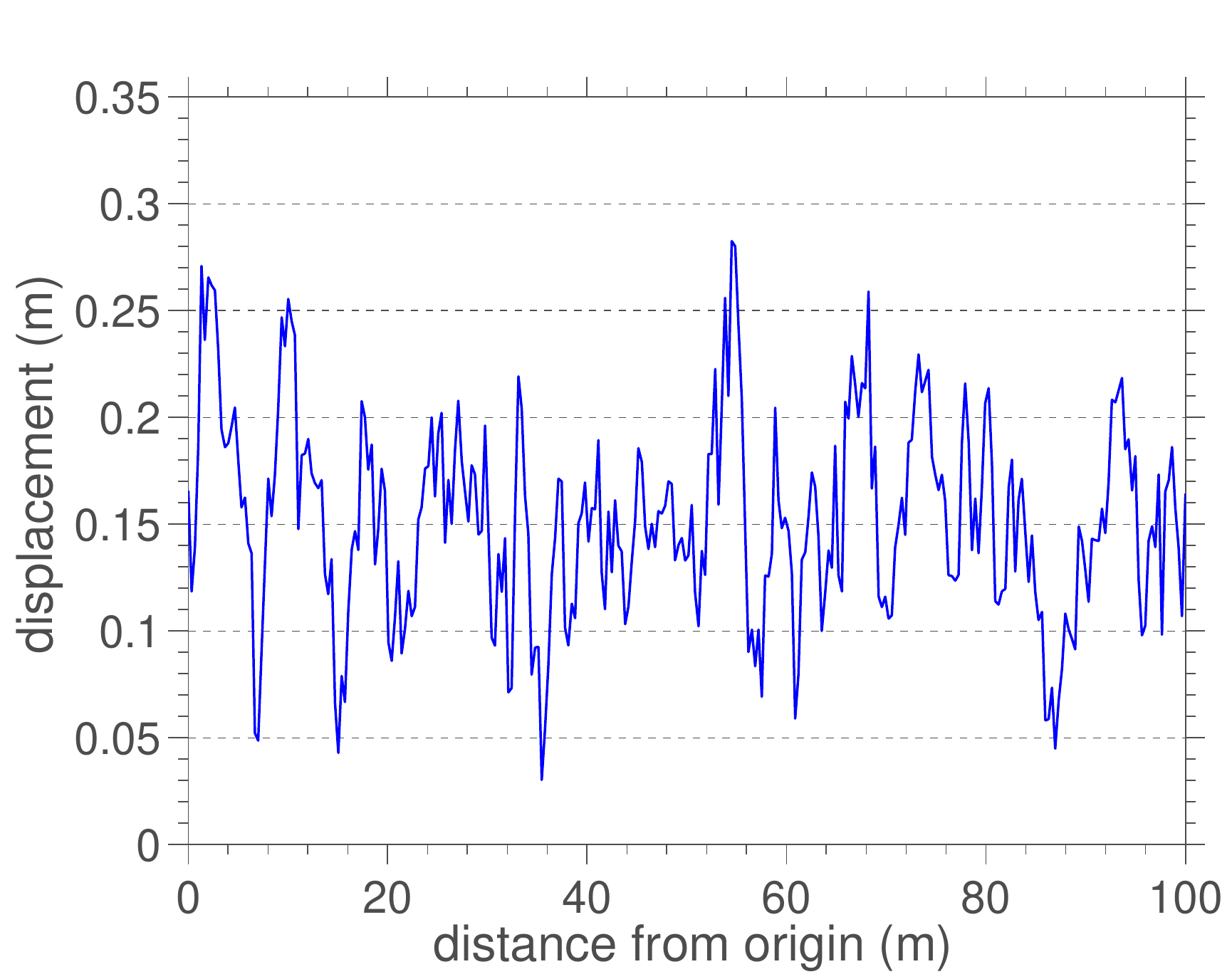}}
	\caption{Illustration of sample paths associated to the stationary
	square-integrable random process generated by the truncated 
	KL decomposition, with $v = 12$ km/h, $a_{corr} = 1$ m, 
	$\stddev{\randvar{y}} = 0.175$ m and $\mean{\randvar{y}} = 0.5$ m.}
	\label{kl_rand_proc_fig}
\end{figure}

The simulations reported here use a family of zero-mean
uncorrelated Gaussian random variables for 
$\left\lbrace \randvar{Y}_n \right\rbrace_{n=1}^{N_{KL}}$, 
which generate a stochastic process which sample paths
can be seen in Figure~\ref{kl_rand_proc_fig}. 
From the qualitative point of view
these realizations of the random process emulate the tracks shown 
in Figure~\ref{tires_path_fig}.

\subsection{Random nonlinear dynamical system}

Due to the randomness of $\randvar{y}_{e1}$ and $\randvar{y}_{e2}$
the mechanical system response becomes aleatory, described by
the real-valued random processes $\randvar{y}_{1}$, $\bbphi_{1}$
and $\bbphi_{2}$.

Therefore, the mechanical system dynamic behavior evolves 
(almost sure) according to the random nonlinear dynamical system 
defined by 

\begin{eqnarray}
\resizebox{0.92\hsize}{!}{$
			\mat{\randvar{M}} \left( \begin{array}{c} 	\ddot{\randvar{y}}_1(t)\\    \ddot{\bbphi}_1(t)\\	\ddot{\bbphi}_2(t)\\\end{array} \right) +
			\mat{\randvar{N}} \left( \begin{array}{c} 	  \dot{\randvar{y}}^2_1(t)\\  \dot{\bbphi}^2_1(t)\\	\dot{\bbphi}^2_2(t)\\\end{array} \right) +
			\mat{\randvar{C}} \left( \begin{array}{c} 	   \dot{\randvar{y}}_1(t)\\      \dot{\bbphi}_1(t)\\	     \dot{\bbphi}_2(t)\\\end{array} \right) +
			\mat{\randvar{K}} \left( \begin{array}{c} 	            \randvar{y}_1(t)\\               \bbphi_1(t)\\	              \bbphi_2(t)\\\end{array} \right) 
			= \randvar{g} - \randvar{h}, ~~~~ a.s.,
			$ }
	\label{stoch_dynamics}
\end{eqnarray}

\noindent
where the real-valued random matrices/vectors $\mat{\randvar{M}}$, 
$\mat{\randvar{N}}$, $\mat{\randvar{C}}$, $\mat{\randvar{K}}$, 
$\randvar{g}$ and $\randvar{h}$ are stochastic versions 
of the matrices/vectors $\mat{M}$, $\mat{N}$, $\mat{C}$, $\mat{K}$,
$\vec{g}$ and $\vec{h}$.

\subsection{Monte Carlo method: the stochastic solver}
\label{MC_method}

Monte Carlo (MC) method \citep{kroese2011,cunhajr2014p1355} 
is employed to compute the propagation of uncertainties of the 
random parameters through the nonlinear dynamics defined by 
Eq.(\ref{stoch_dynamics}). The convergence of MC simulations 
is evaluated through the map 
$\texttt{conv}: n_s \in \N \mapsto \texttt{conv} (n_{s}) \in \R $, 
where $n_{s}$ is the number of MC realizations, $\SSpt_n$ denotes 
the n-the MC realization, $[t_0,t_f]$ is the time interval of analysis, and

\begin{equation}
		\texttt{conv}(n_{s}) = 
		\left(
		\frac{1}{n_{s}} \sum_{n=1}^{n_{s}} 
		\int_{t=t_0}^{t_f} \left( \randvar{y}_1(t,\SSpt_n)^2 + \bbphi_1(t,\SSpt_n)^2 + \bbphi_2(t,\SSpt_n)^2 \right)\,dt
		\right)^{1/2}.
		\label{MC_conv_eq}
\end{equation}

\noindent
This metric allows one to evaluate the convergence of the 
approximation $\transp{\left( \randvar{y}_1(t,\SSpt_n), \bbphi_1(t,\SSpt_n), \bbphi_2(t,\SSpt_n) \right)}$ 
in the mean-square sense. See \cite{soize2005p623} for further details.



\section{Numerical experiments}
\label{num_results}

The physical parameters adopted in the simulation of the mechanical system
are presented in Table~\ref{physical_param_tab}. They correspond to the nominal
parameters of an sprayer tower model Arbus Multisprayer 4000, illustrated in 
Figure~\ref{mech_sys_fig}, whose values can be seen in \cite{sartoriJunior2009p417}.

\begin{table}[h!]
	\centering
	\caption{Physical parameters for the mechanical system used in the simulations.}
	\vspace{5mm}
	\begin{tabular}{ccl}
		\toprule
		parameter & value & unit  \\
		\midrule
		$m_1$ & $6500$     & kg \\
		$m_2$ & $800$       & kg \\
		$L_1$  & $0.2$        & m \\
		$L_2$  & $2.4$        & m \\
		$I_1$   & $6850$     & kg \, m$^2$ \\
		$I_2$   & $6250$     & kg \, m$^2$ \\
		$k_1$  & $465 \times 10^{3}$ & N/m \\
		$k_2$  & $465 \times 10^{3}$ & N/m \\
		$c_1$  & $5.6 \times 10^{3}$     & N/m/s \\
		$c_2$  & $5.6 \times 10^{3}$     & N/m/s \\		
		$B_1$  & $0.85$ & m \\
		$B_2$  & $0.85$ & m \\
		$k_T$  & $100 \times 10^{3}$       & N/rad \\
		$c_T$  & $~40 \times 10^{3}$       & N m/rad/s \\
		\bottomrule
	\end{tabular}
	\label{physical_param_tab}
\end{table}

Moreover, the parameters which define the random loadings can be seen in 
Table~\ref{stochastic_param_tab}. They are obtained via educated judgment,
trial and error, always checking if the behavior of the tower sprayer was in 
agreement with the intuition of the authors about this physical system. In fact,
it is reasonable to assume that the radius of mutual influence (correlation)
between soil irregularities has the same order of magnitude as the sprayer tower tires 
diameters. Once each tire has a diameter of the order of magnitude of $1\,m$, 
it is assumed that $a_{corr} = 1\,m$. The displacements $\randvar{y}_{e1}(t)$ 
and $\randvar{y}_{e2}(t)$ correspond to vertical translations of the tires 
centroids, which on a soil without irregularities will be approximately 
$0.5$ m above the ground (half of the tire diameter). Therefore,
$\mean{y} = 0.5$ m is adopted.  The choice of standard deviation values corresponds 
to a dispersion level of 35\%, which provides stringent soil-irregularities 
induced loadings. The latter is necessary to investigate severe conditions of 
lateral (horizontal) vibrations.

\begin{table}[h!]
	\centering
	\caption{Parameters that define the stochastic loadings.}
	\vspace{5mm}
	\begin{tabular}{ccl}
		\toprule
		parameter & value & unit  \\
		\midrule
		$N_{KL}$               & $403$    & ---\\
		$a_{corr}$             & $1$        & m\\
		$\stddev{y_{e1}}$ & $0.175$ & m \\
		$\stddev{y_{e2}}$ & $0.175$ & m \\
		$\mean{y_{e1}}$   & $0.5$     & m \\
		$\mean{y_{e2}}$   & $0.5$     & m \\
		            $v$           & $12$                                 & km/h \\
		\bottomrule
	\end{tabular}
	\label{stochastic_param_tab}
\end{table}

A representative band of frequencies  for the present problem is given by
$\mathcal{B} = [0,5]$ Hz, once the sprayer tower operates on the low 
frequency range. Thus, the evolution of the nonlinear dynamic system is 
addressed using a nominal time step $\Delta t = 1 \times 10^{-3}$ s, 
which is refined whenever necessary to capture the nonlinear effects.

\subsection{Nonlinear dynamics animation}

In a first moment, the nonlinear dynamics is explored in the
temporal window defined by $[t_0,t_f]=[0,30]$ s. This time-interval
corresponds to a traveled path of $100$ m, such as those shown 
in Figure~\ref{tires_path_fig}.

\begin{figure}[ht!]
	\centering
	\subfigure[$t=0$s  ]{\includegraphics[scale=0.28]{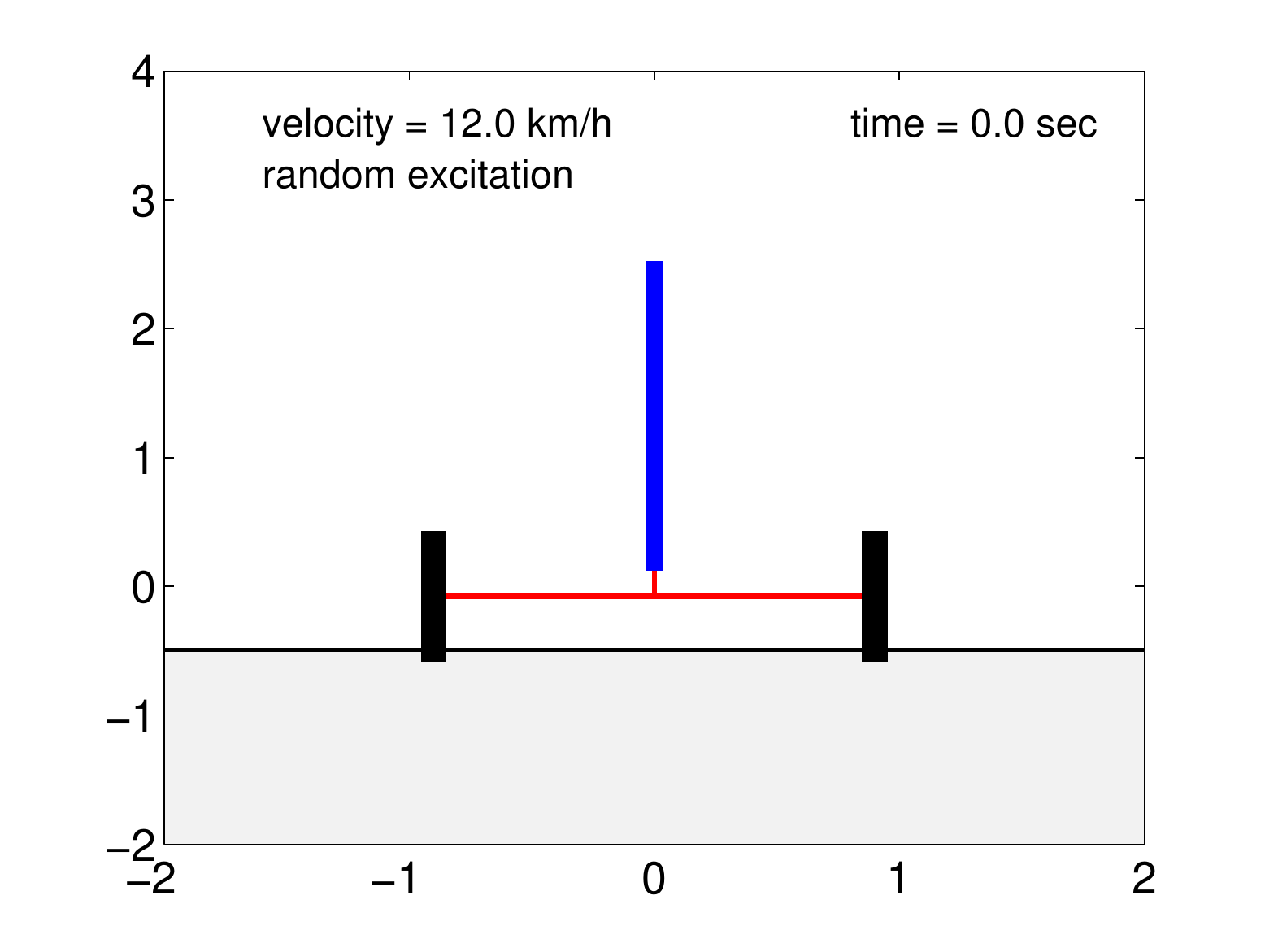}}
	\subfigure[$t=3$ s  ]{\includegraphics[scale=0.28]{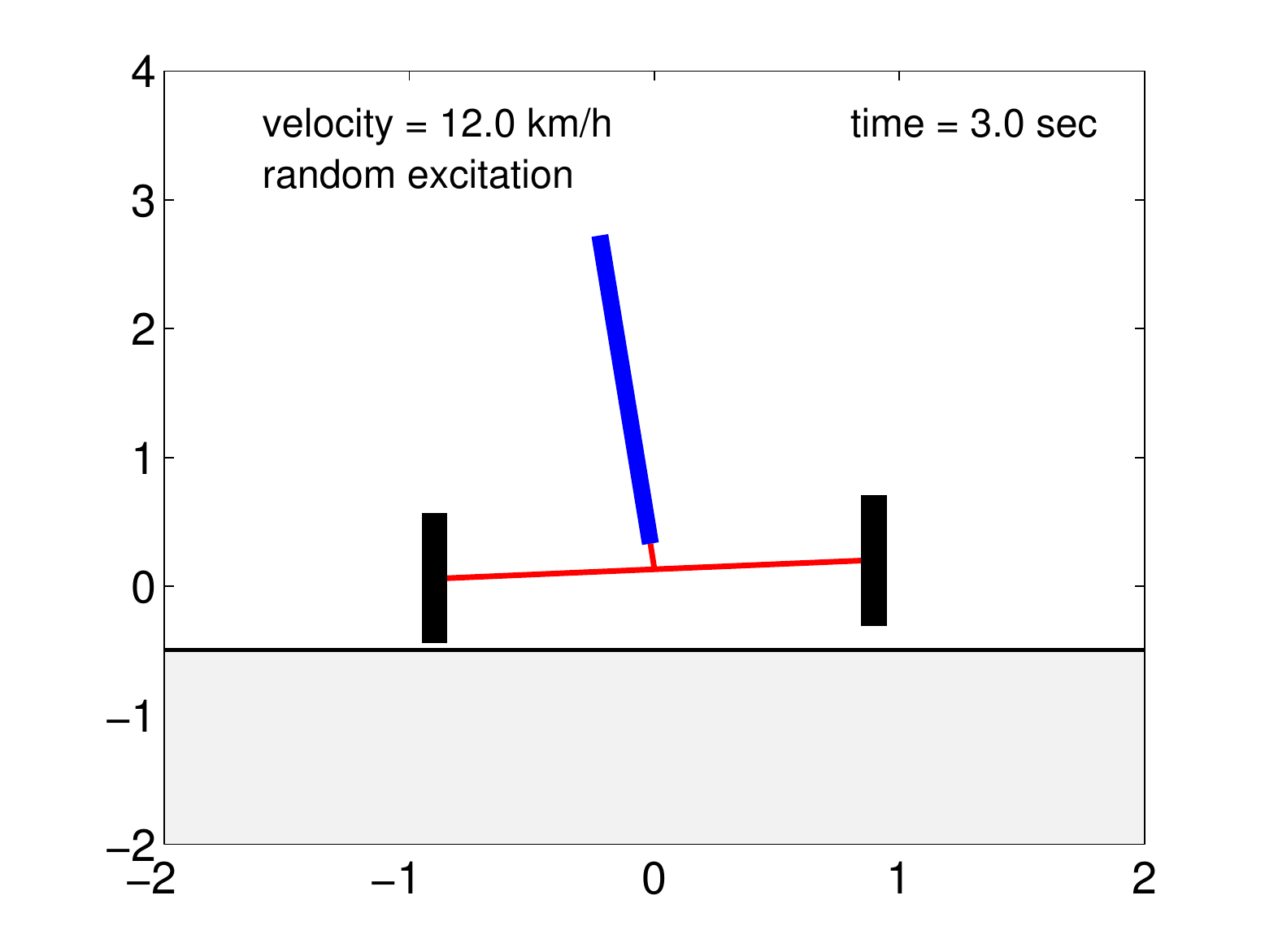}}
	\subfigure[$t=7$ s]{\includegraphics[scale=0.28]{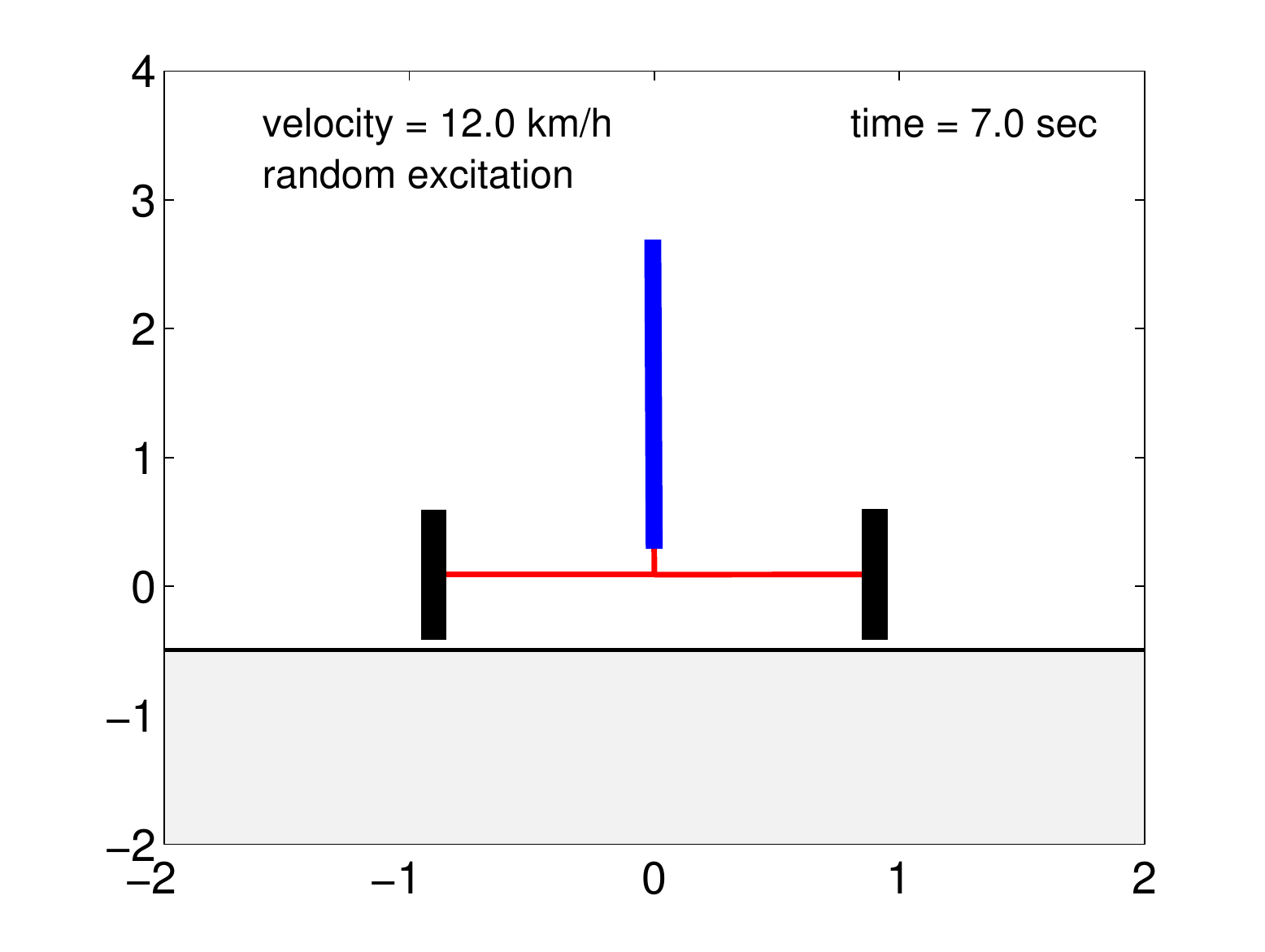}}\\
	\subfigure[$t=11$ s]{\includegraphics[scale=0.28]{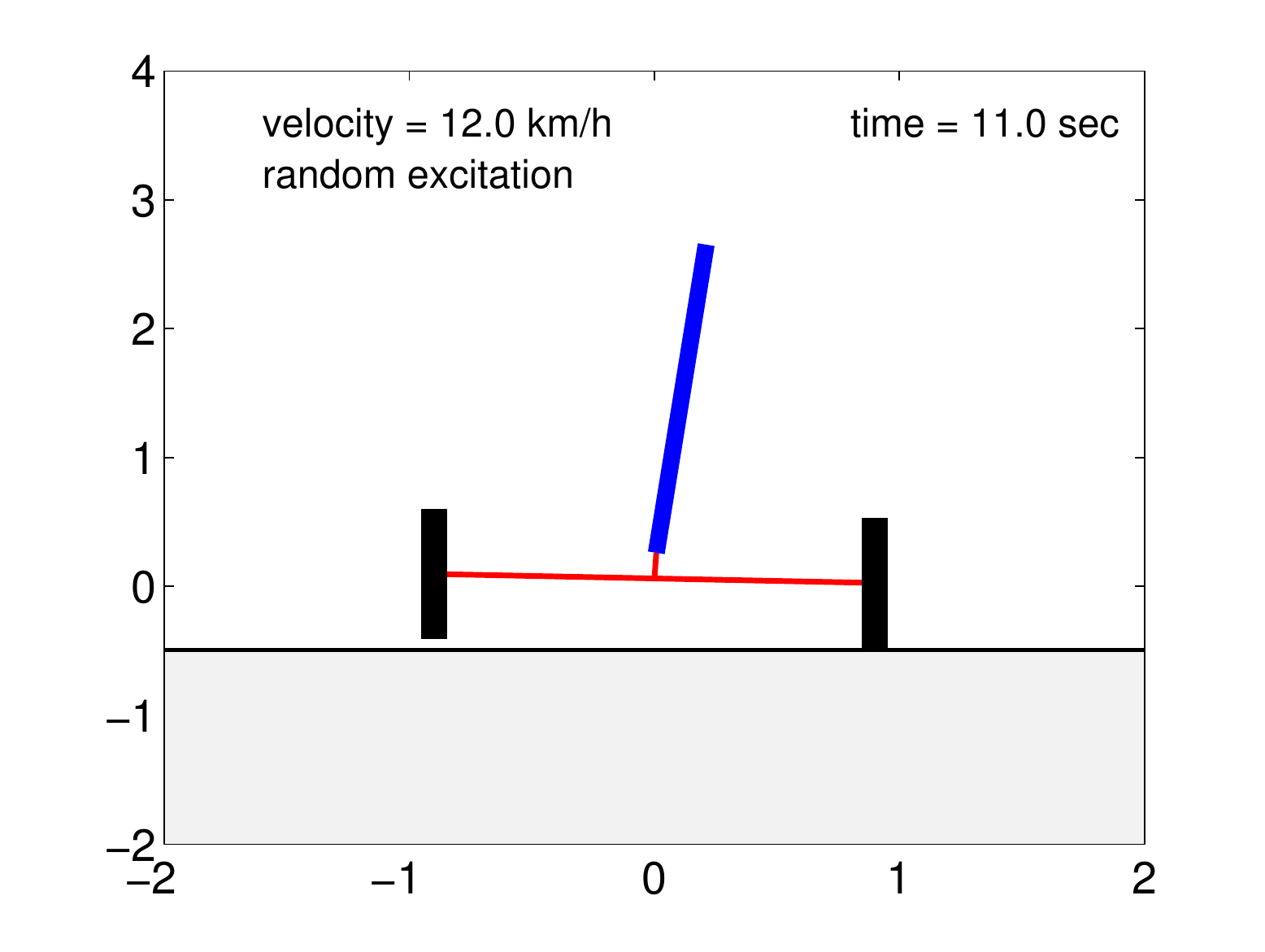}}
	\subfigure[$t=15$ s]{\includegraphics[scale=0.28]{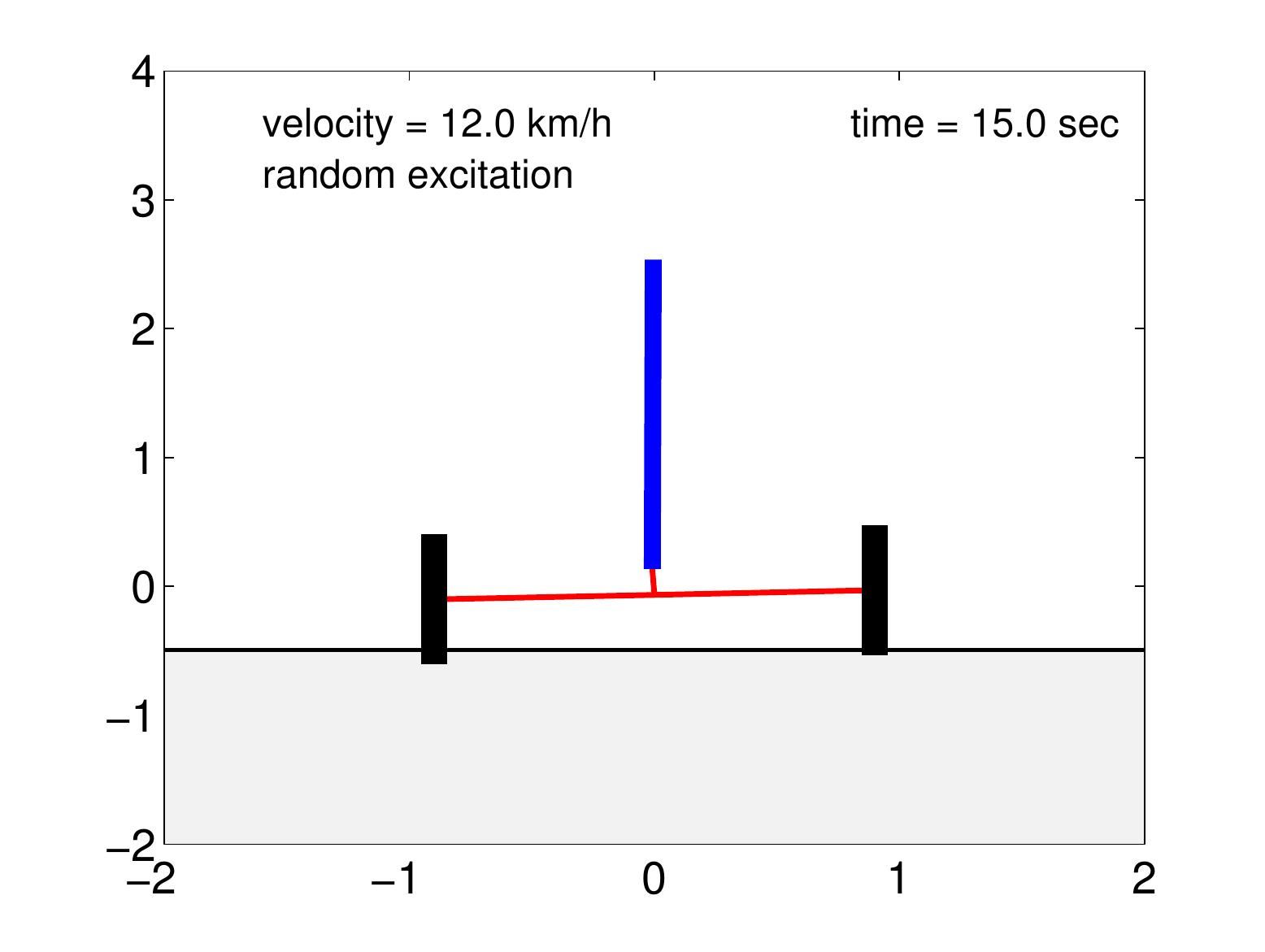}}
	\subfigure[$t=18$ s]{\includegraphics[scale=0.28]{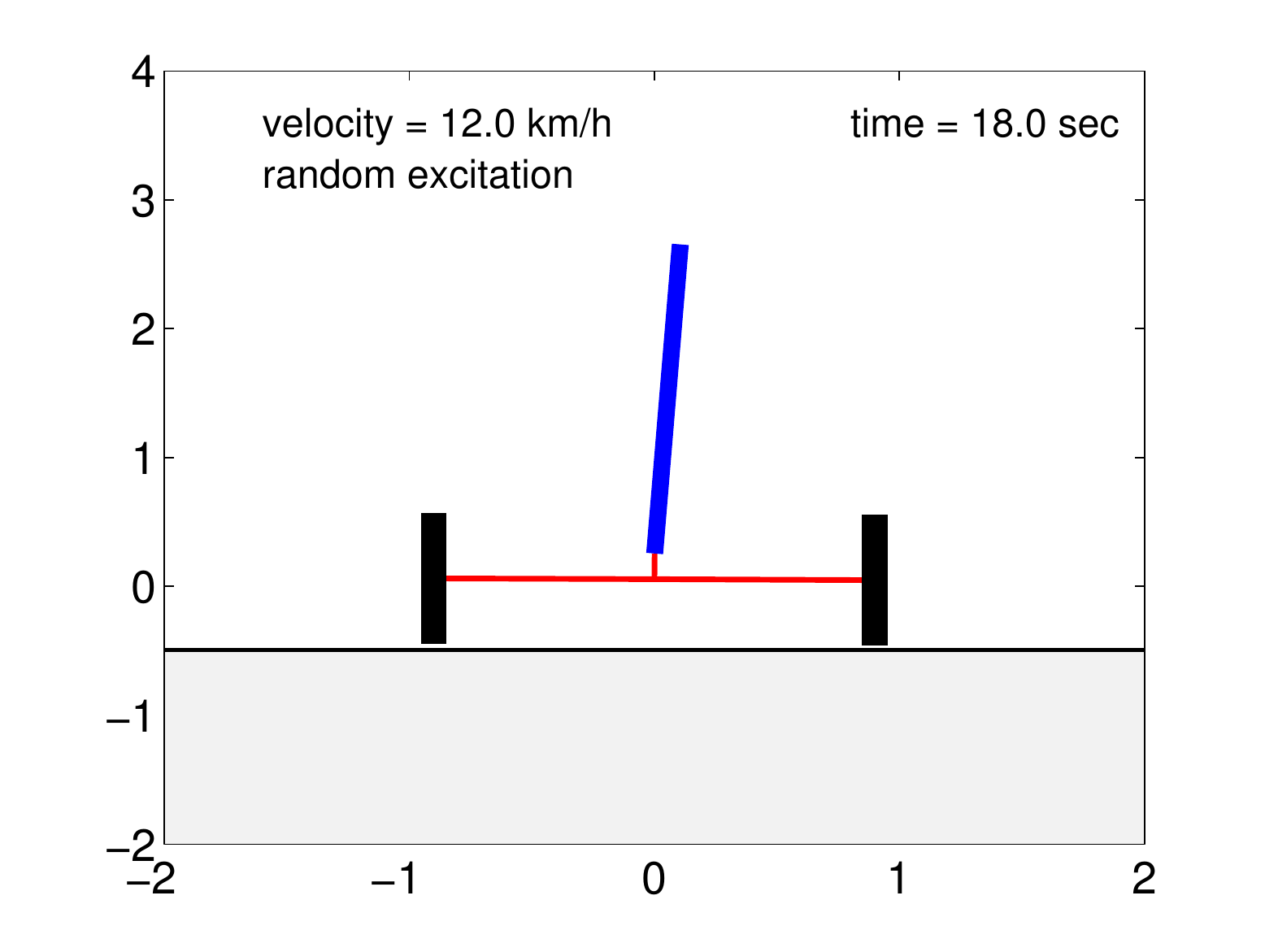}}\\
	\subfigure[$t=22$ s]{\includegraphics[scale=0.28 ]{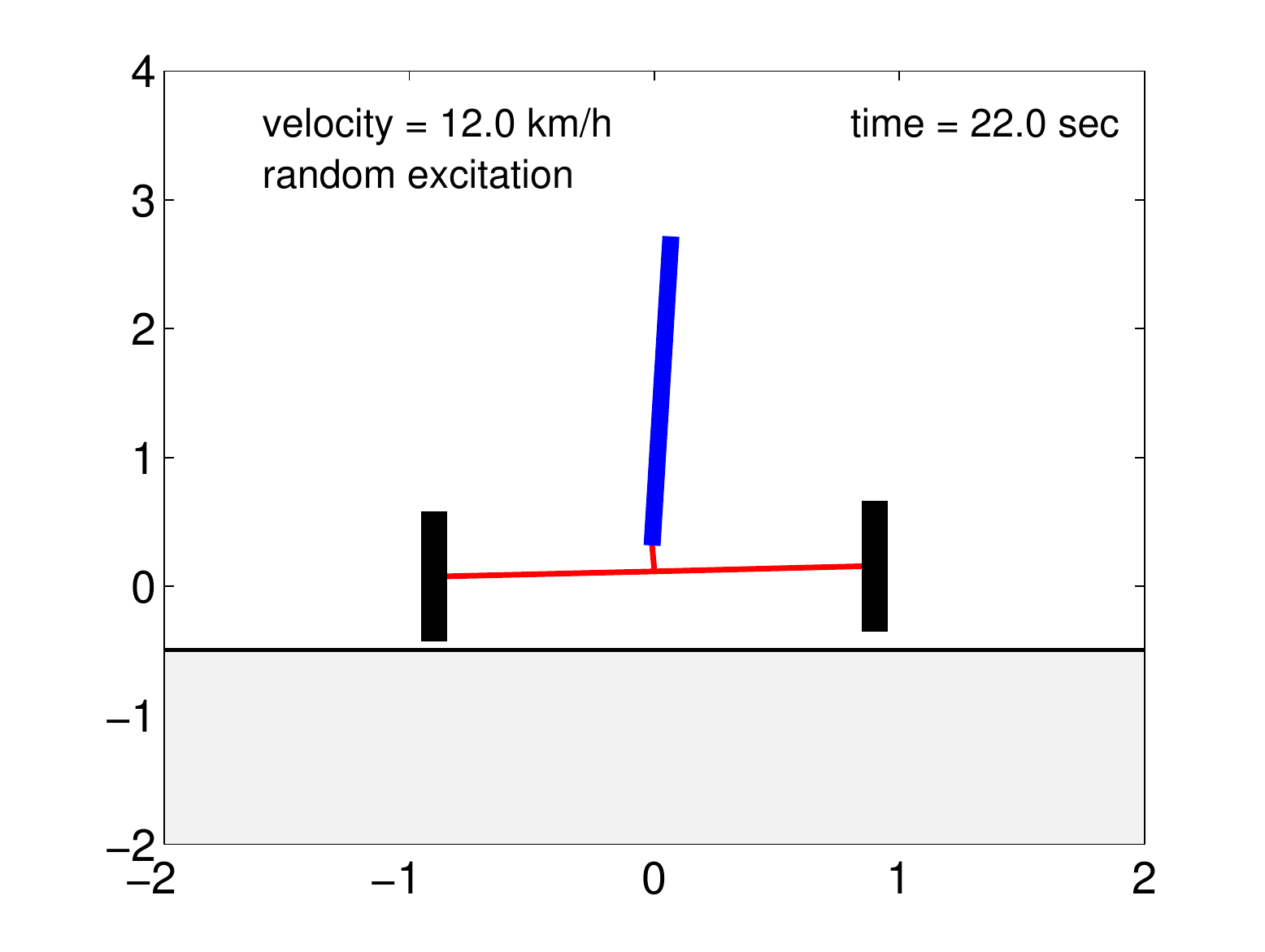}}
	\subfigure[$t=26$ s]{\includegraphics[scale=0.28]{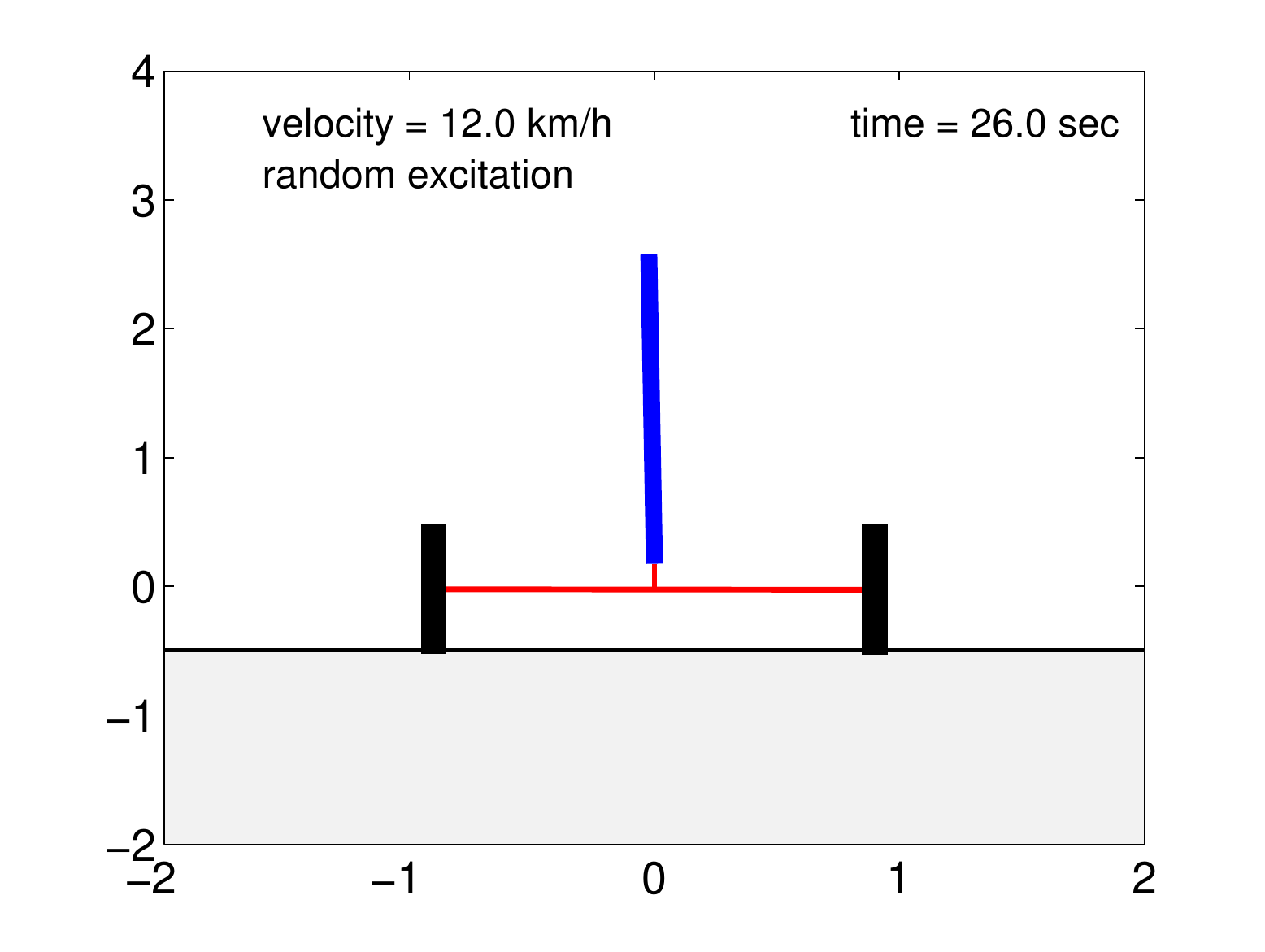}}
	\subfigure[$t=30$ s]{\includegraphics[scale=0.28]{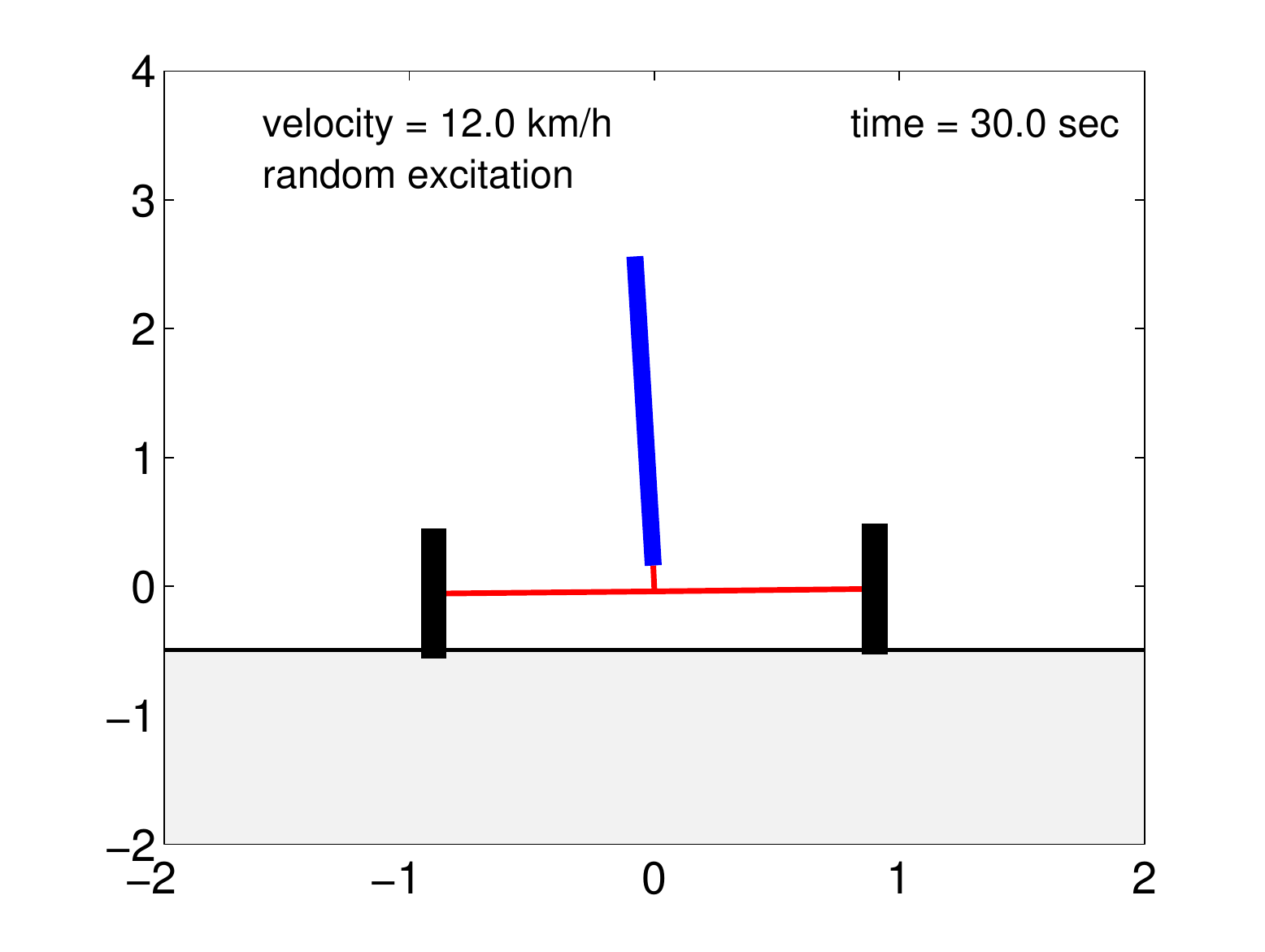}}
	\caption{Animation of the mechanical system at different instants of time.}
	\label{animation_fig}
\end{figure}

An animation of the mechanical dynamic system, for different instants 
of time in $[t_0,t_f]$, is shown in Figure~\ref{animation_fig}.
In this animation the mechanical system is supported on the ground
(gray shaded region bounded by a black thick line), the tires are 
represented by black vertical rectangles, the red lines correspond
to the trailer and the tower is illustrated as a thicker blue line.
The video animation is available in Supplementary Material~1
\cite{video1}.


\subsection{Time domain analysis}

\begin{figure}[h!]
	\centering
	\subfigure[]{\label{tseries_y1_y_2_figA} \includegraphics[scale=0.35]{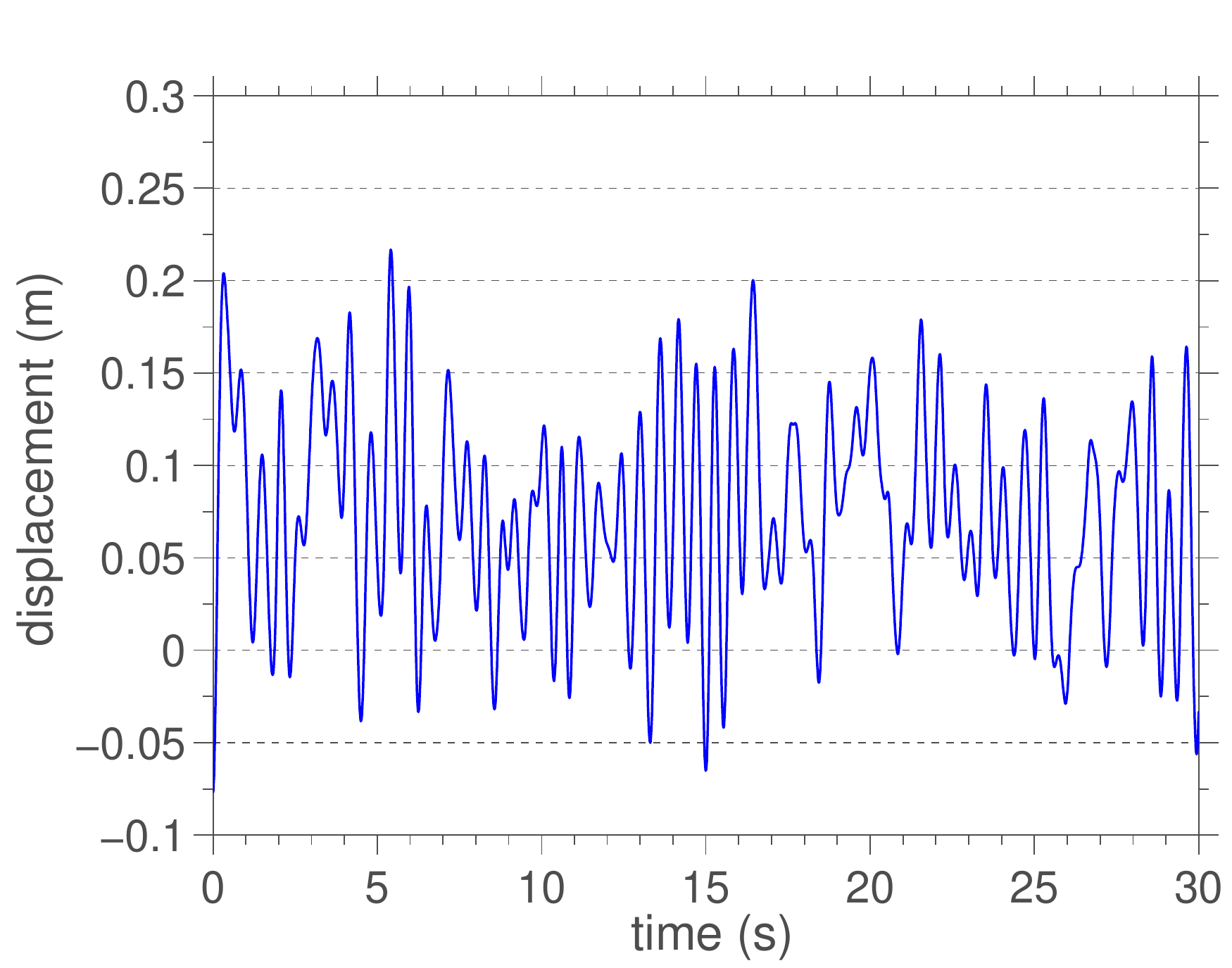}}
	\subfigure[]{	\label{tseries_y1_y_2_figB} \includegraphics[scale=0.35]{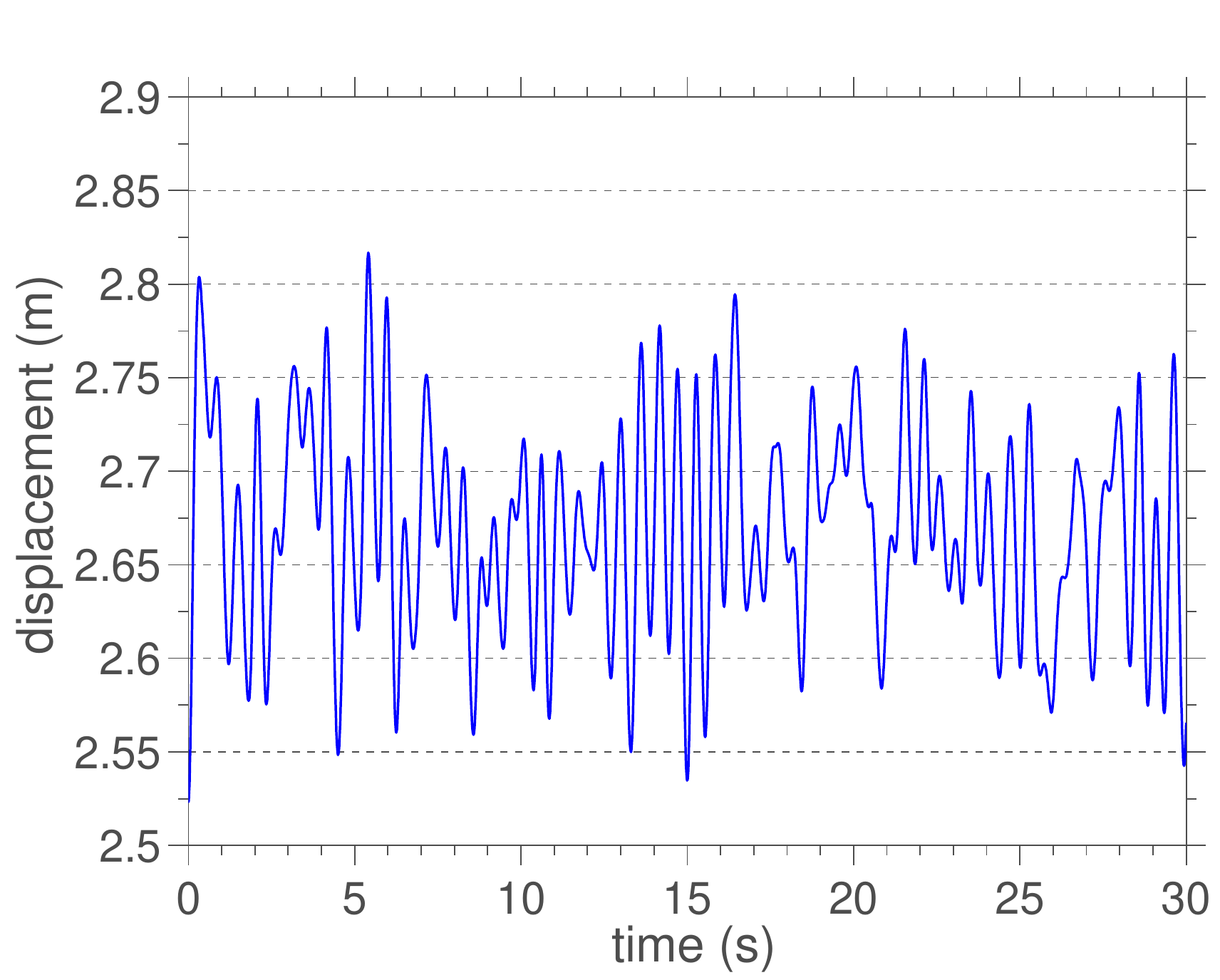}}\\
	\subfigure[]{	\label{tseries_y1_y_2_figC} \includegraphics[scale=0.35]{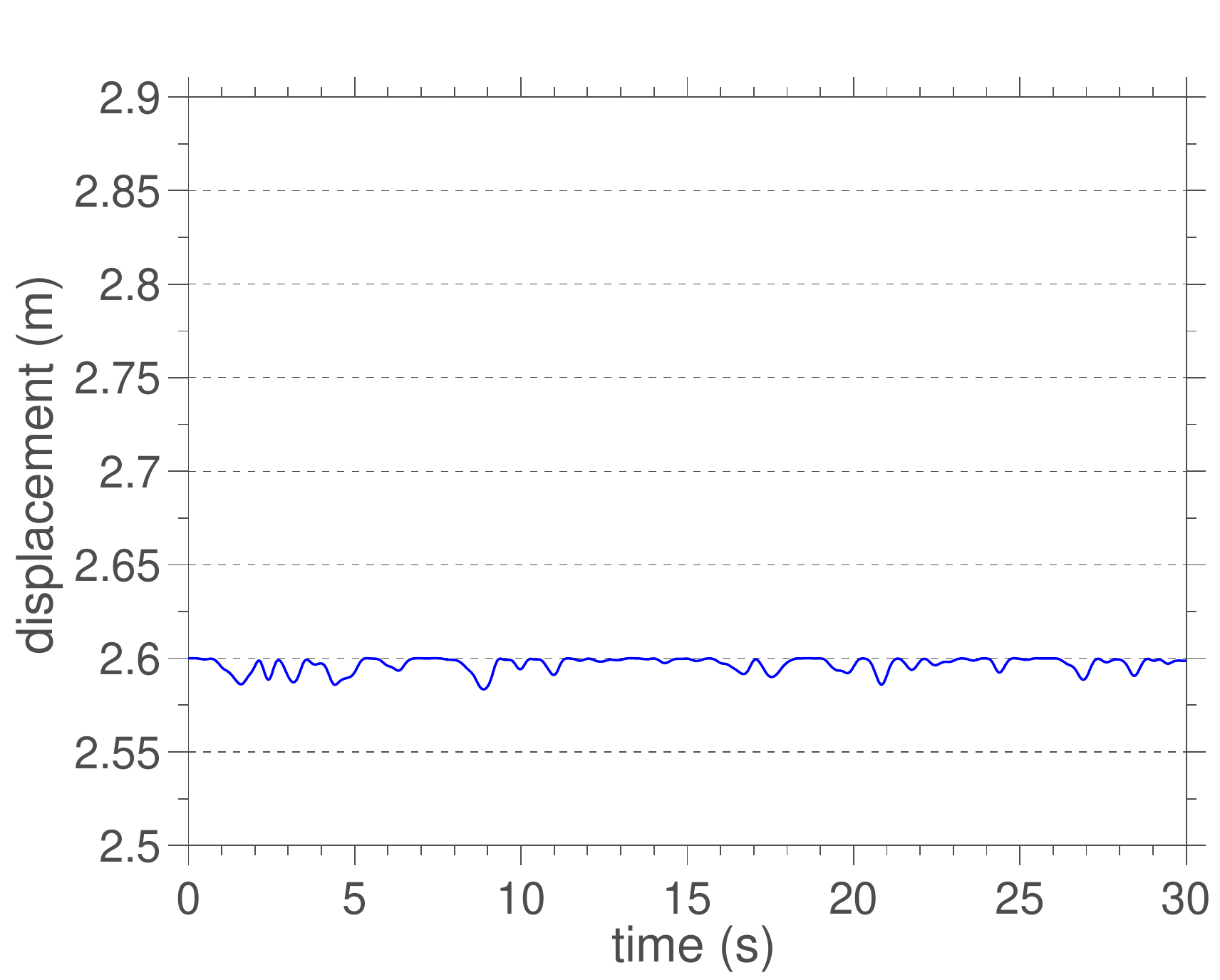}}
	\caption{Time series of vertical nonlinear dynamics. (a) trailer displacement $y_1$;
	(b) tower displacement $y_2$; (b) difference between $y_2$ and $y_1$.}
	\label{tseries_y1_y_2_fig}
\end{figure}

\begin{figure}[h!]
	\centering
	\subfigure[]{\includegraphics[scale=0.34]{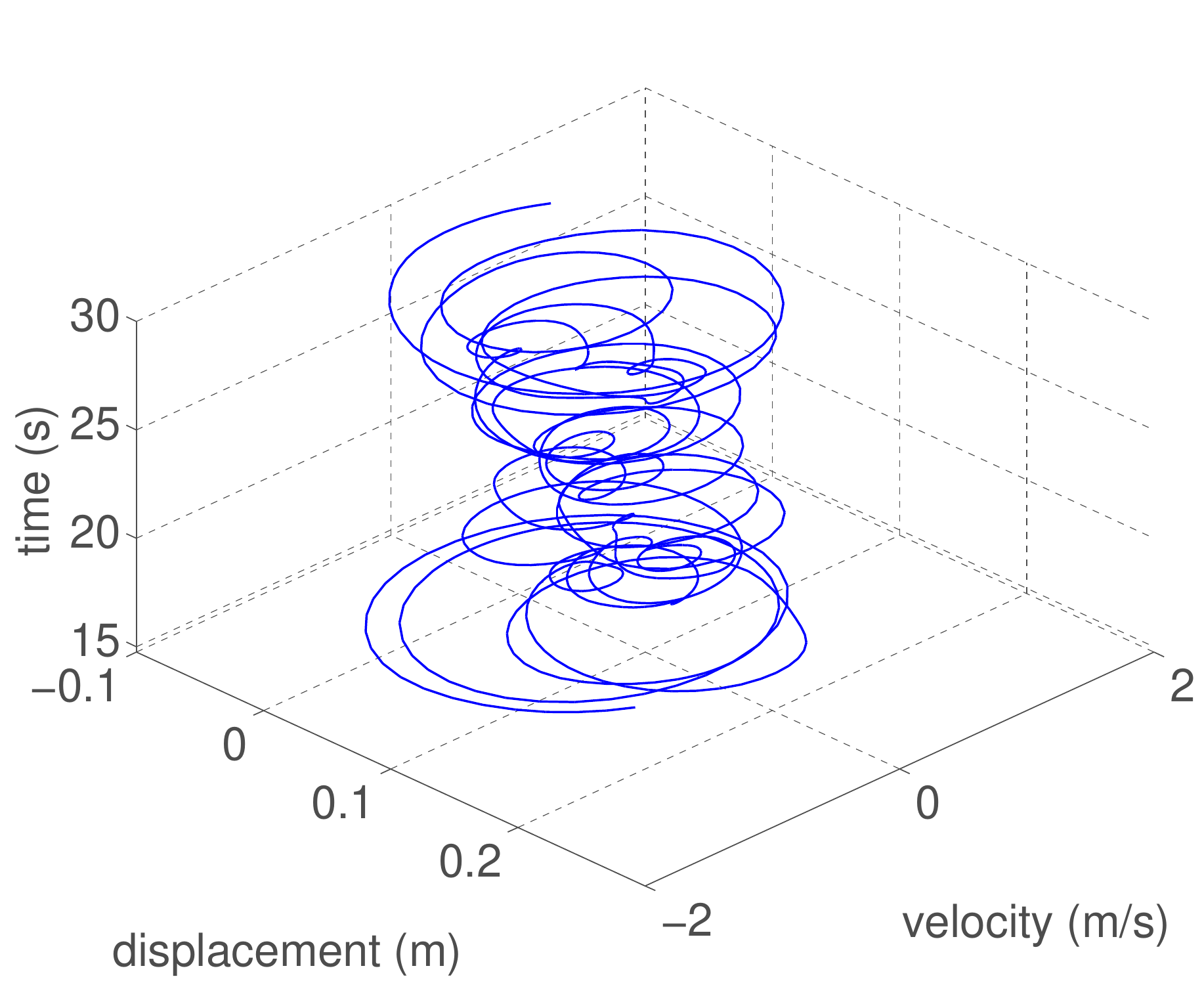}}
	\subfigure[]{	\includegraphics[scale=0.35]{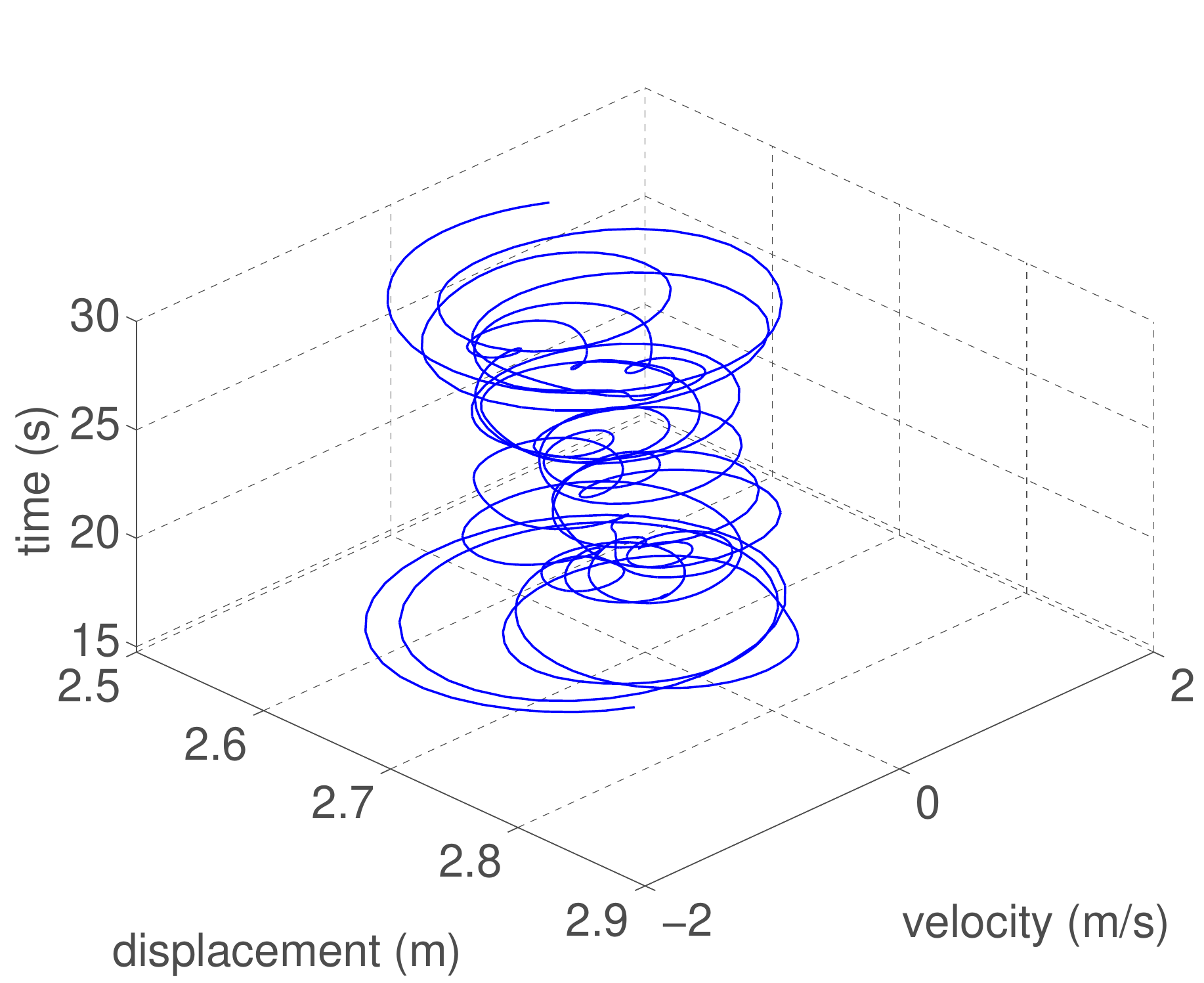}}\\
	\subfigure[]{	\includegraphics[scale=0.34]{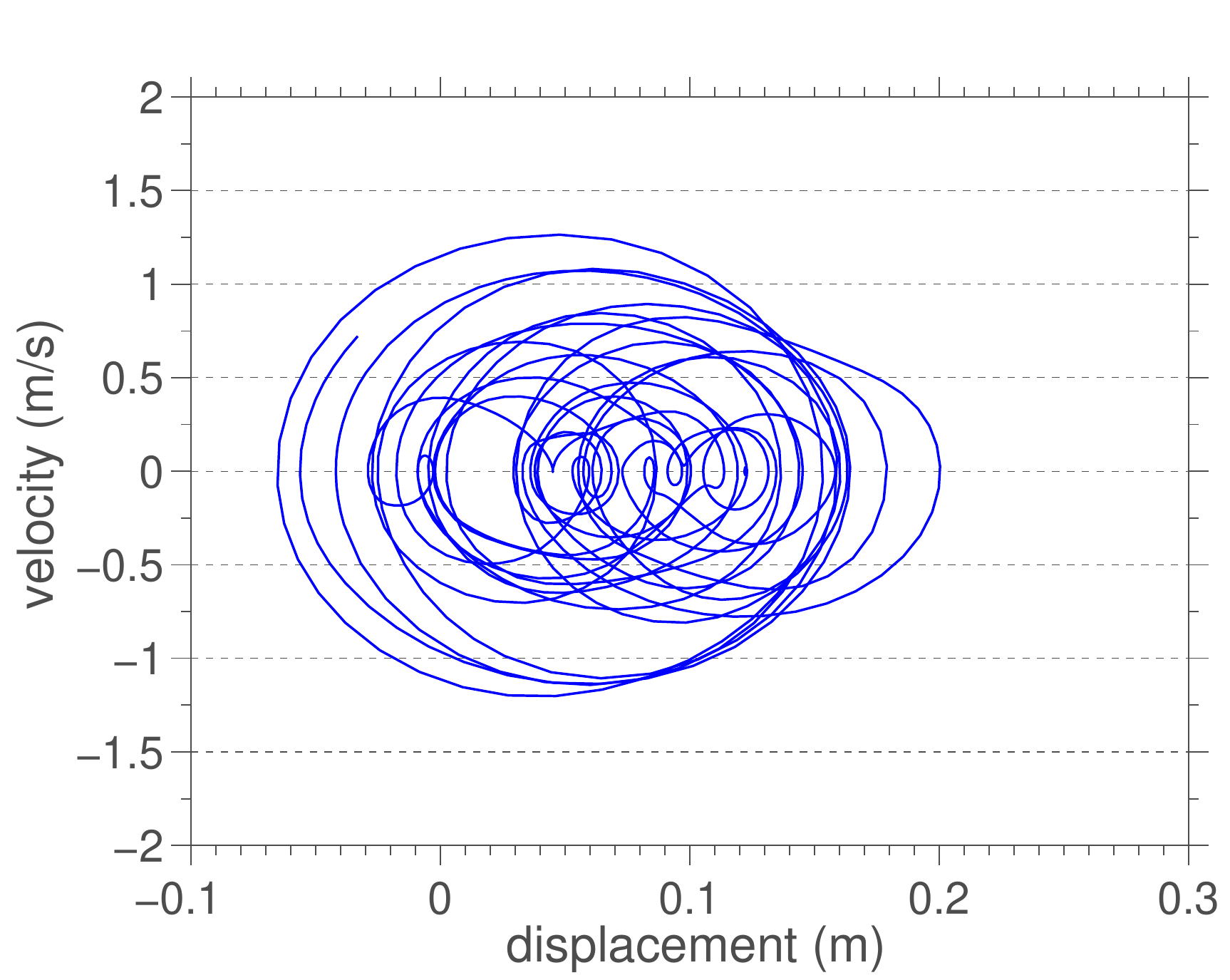}}
	\subfigure[]{	\includegraphics[scale=0.35]{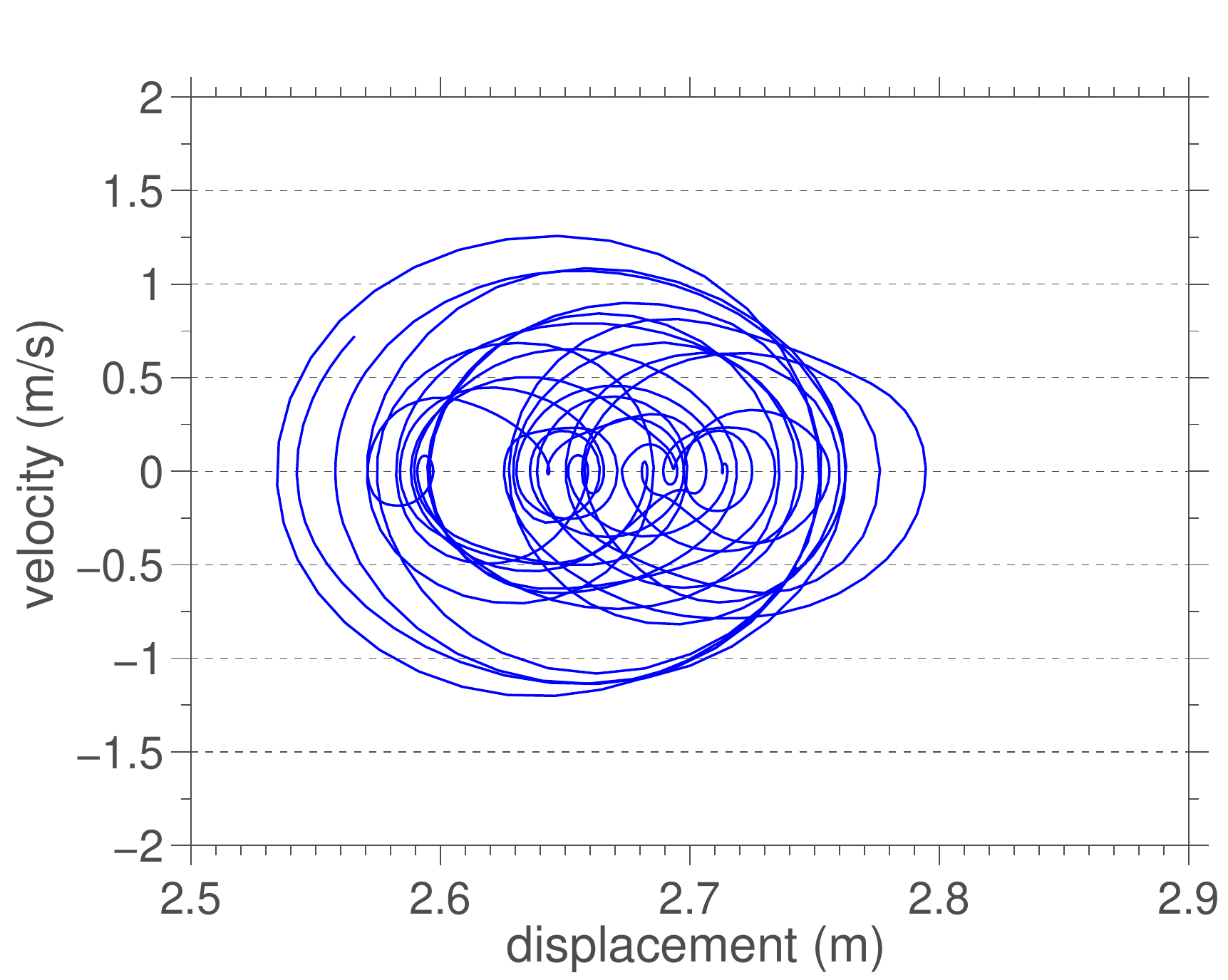}}
	\caption{Projections of vertical dynamics phase space trajectories.
	(a) $y_1$ atractor in $\R^3$; (b) $y_2$ atractor in $\R^3$;
	(c) $y_1$ atractor in $\R^2$; (d) $y_2$ atractor in $\R^2$.}
	\label{phase_space_y1_y2_fig}
\end{figure}

The time series corresponding to the trailer/tower vertical 
dynamics $y_1$/$y_2$ can be seen in Figure~\ref{tseries_y1_y_2_fig},
while the corresponding phase space trajectory projections 
(in $\R^3$ and $\R^2$) are presented in Figure~\ref{phase_space_y1_y2_fig}.

It may be noted from Figure~\ref{tseries_y1_y_2_fig} that both $y_1$ and $y_2$ 
have irregular oscillatory behavior, which are quite similar. The difference between 
$y_2$ and $y_1$ is very small, and can be seen in Figure~\ref{tseries_y1_y_2_figC}.
The strong correlation between the two time series is visually noticeable. The trajectories 
projections shown in Figure~\ref{phase_space_y1_y2_fig} corroborate the previous statement.

\begin{figure}[ht!]
	\centering
	\subfigure[]{\includegraphics[scale=0.35]{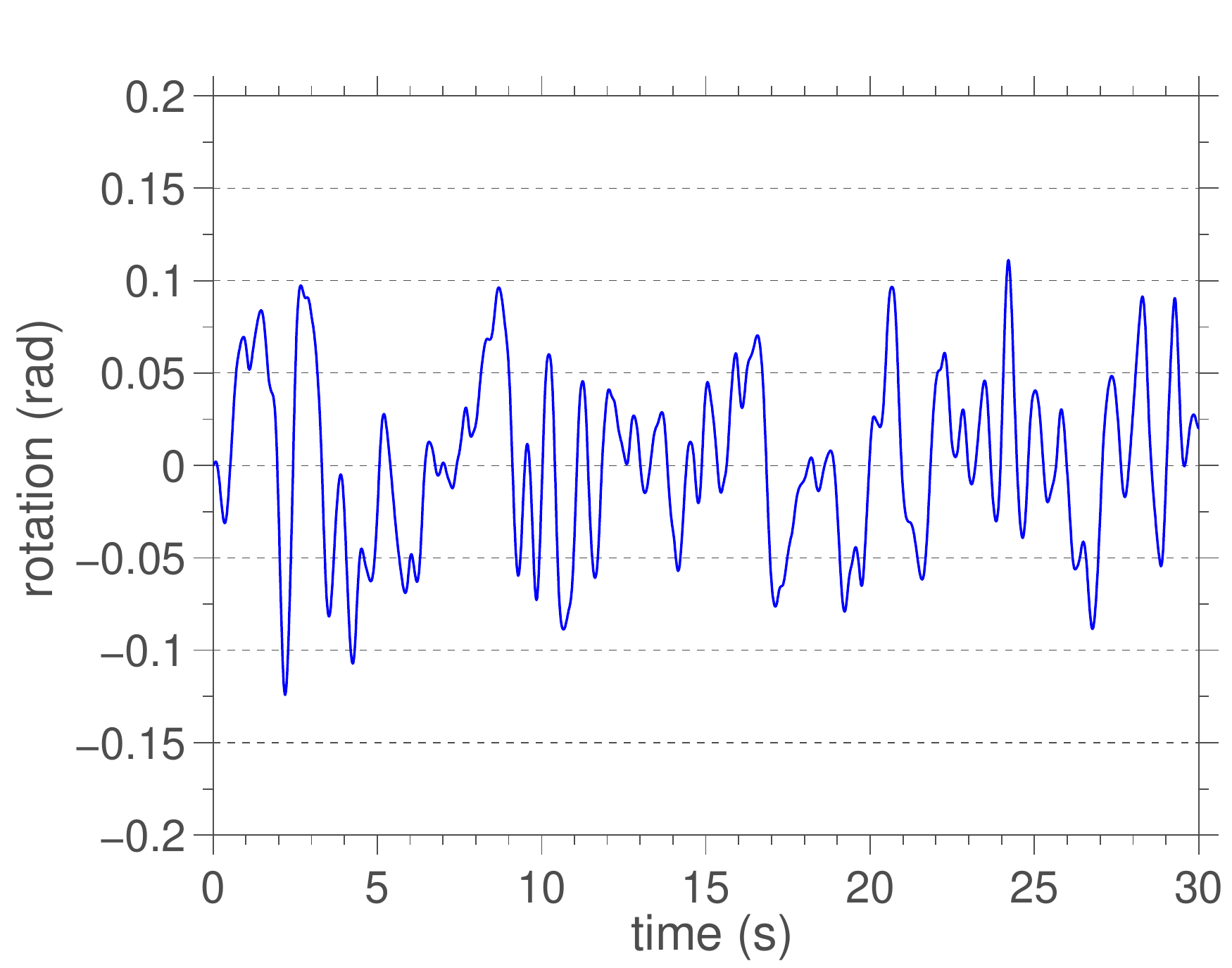}}
	\subfigure[]{	\includegraphics[scale=0.35]{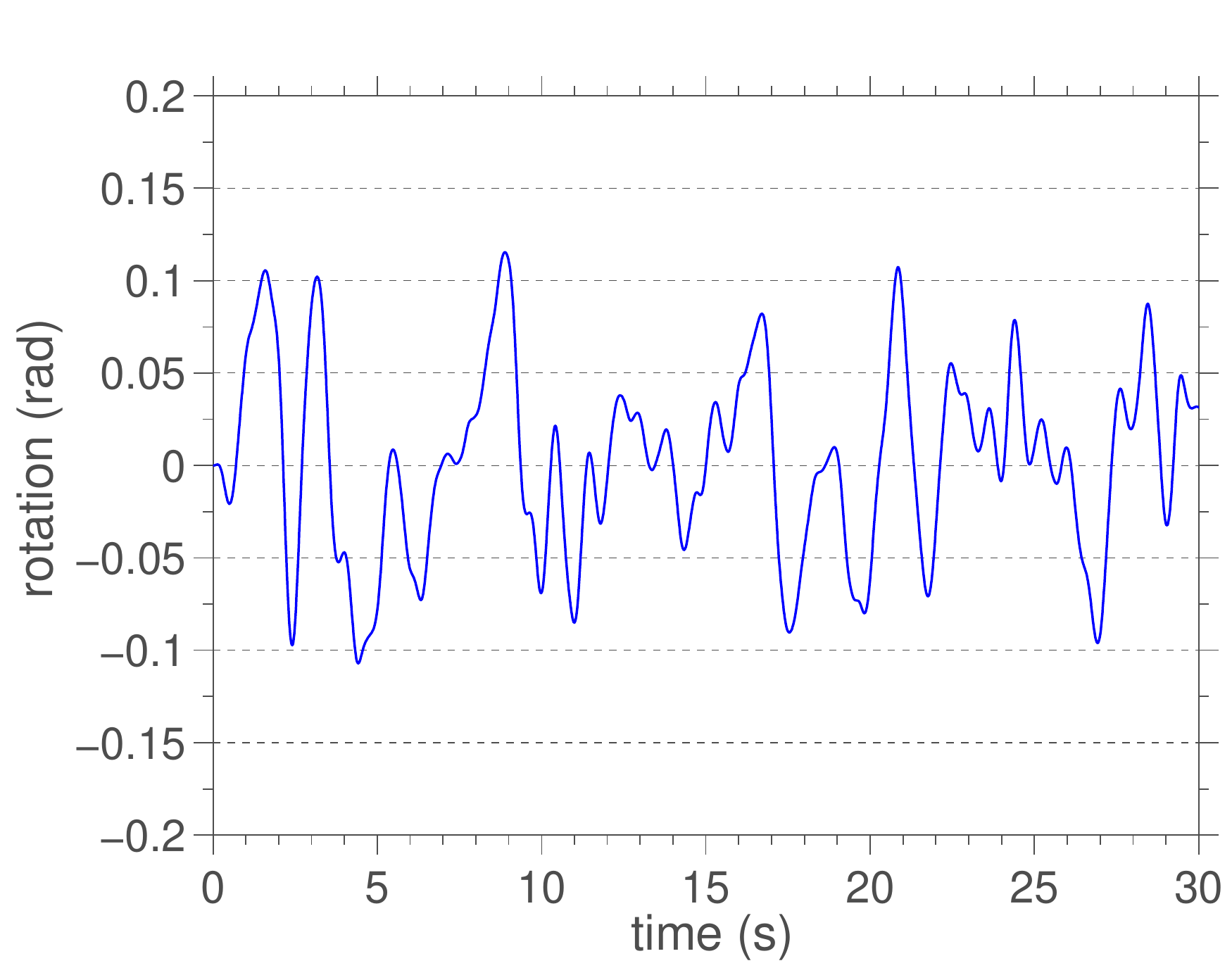}}\\
	\subfigure[]{	\includegraphics[scale=0.35]{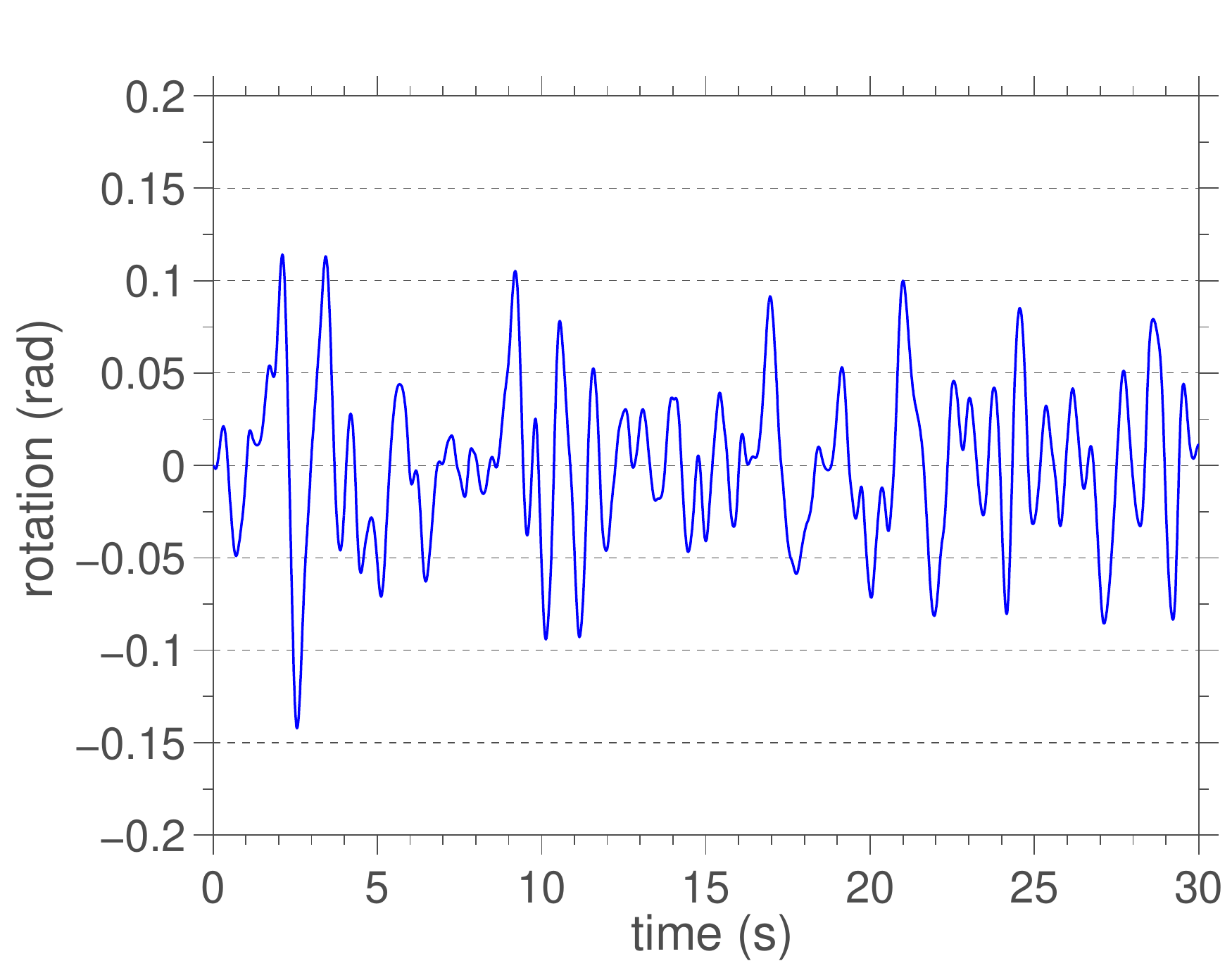}}
	\caption{Time series of rotational nonlinear dynamics. (a) trailer rotation $\phi_1$; (b) tower rotation $\phi_2$;
	(c) difference between $\phi_2$ and $\phi_1$.}
	\label{tseries_phi1_phi2_fig}
\end{figure}

\begin{figure}[ht!]
	\centering
	\subfigure[]{	\includegraphics[scale=0.33]{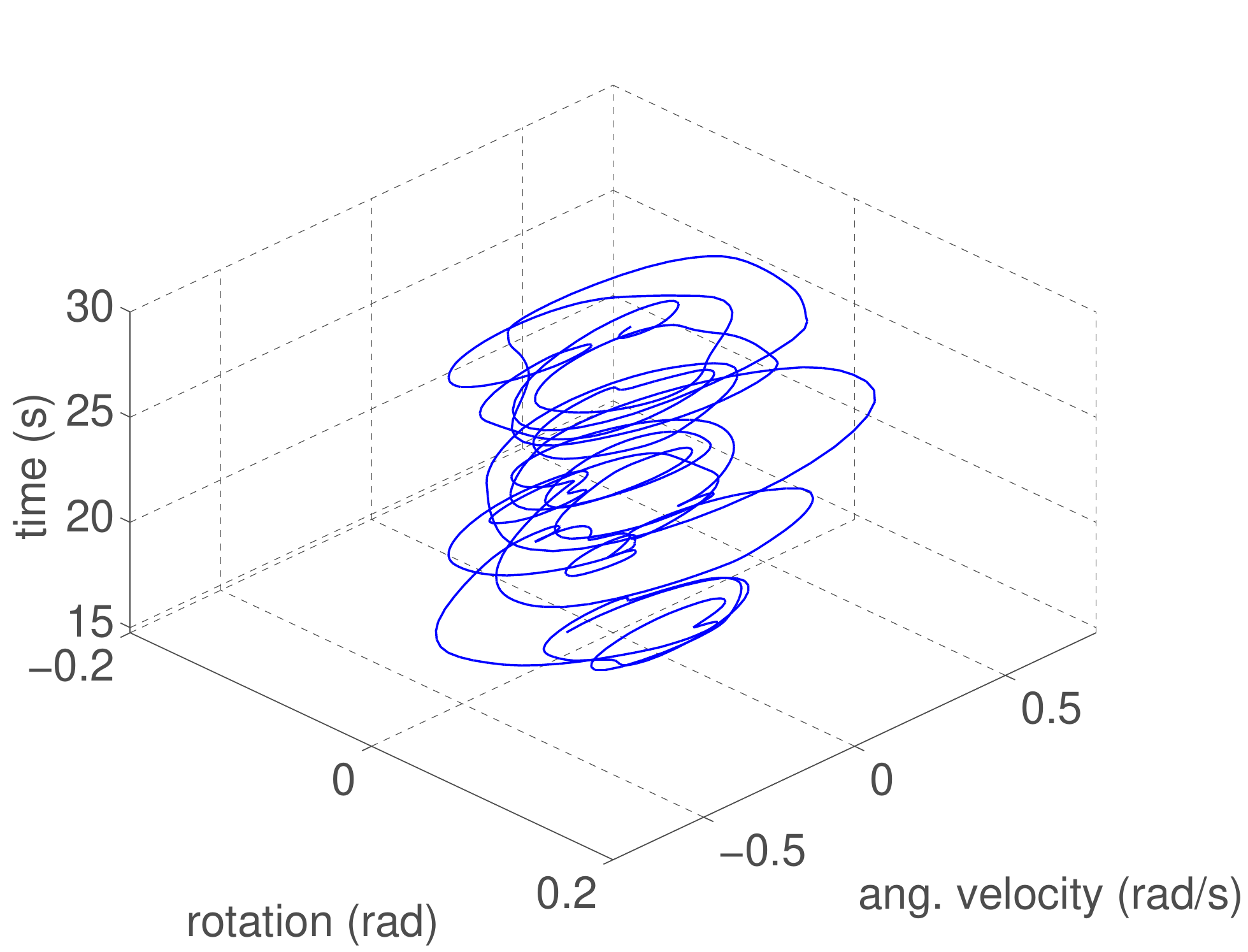}}
	\subfigure[]{	\includegraphics[scale=0.33]{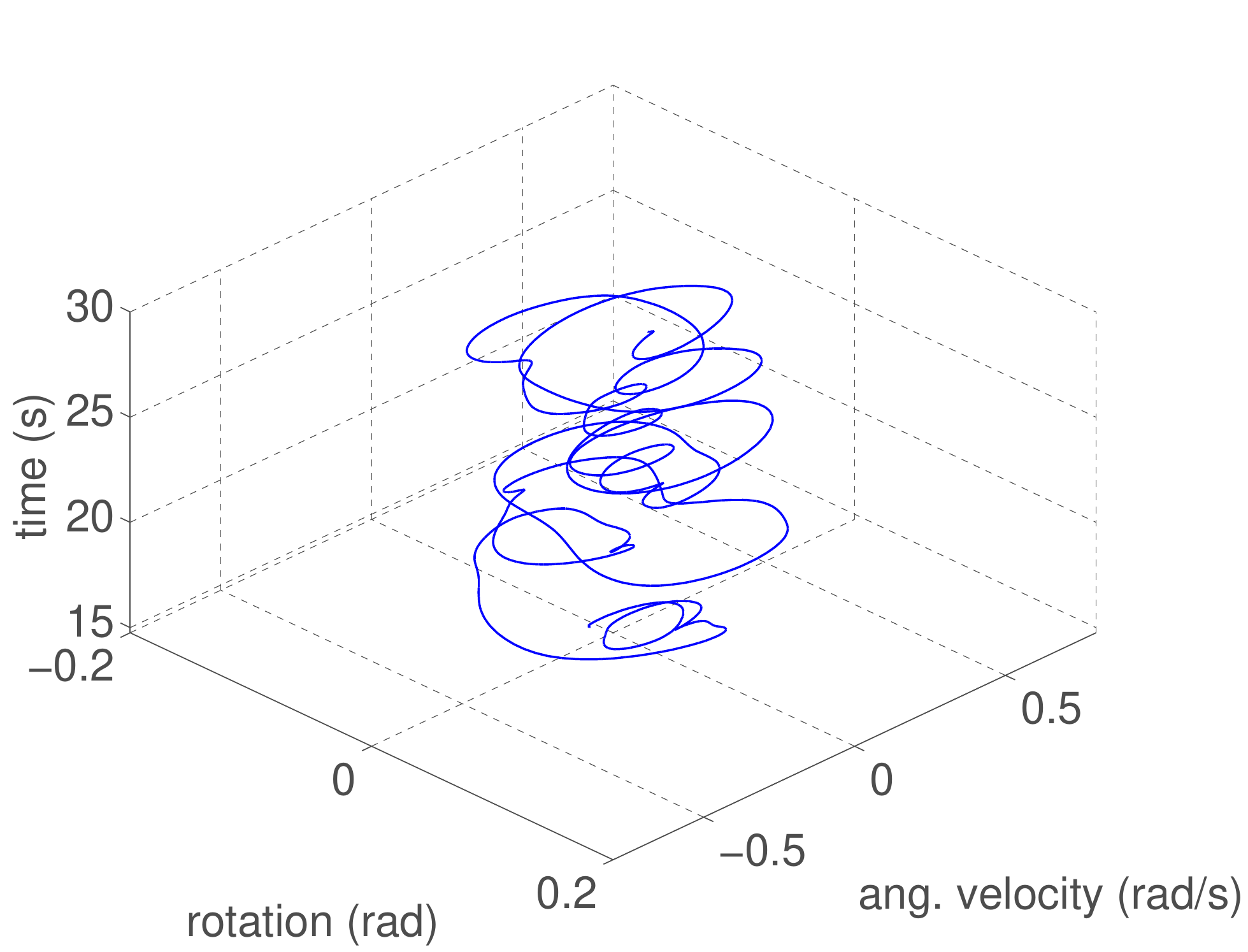}}\\
	\subfigure[]{	\includegraphics[scale=0.34]{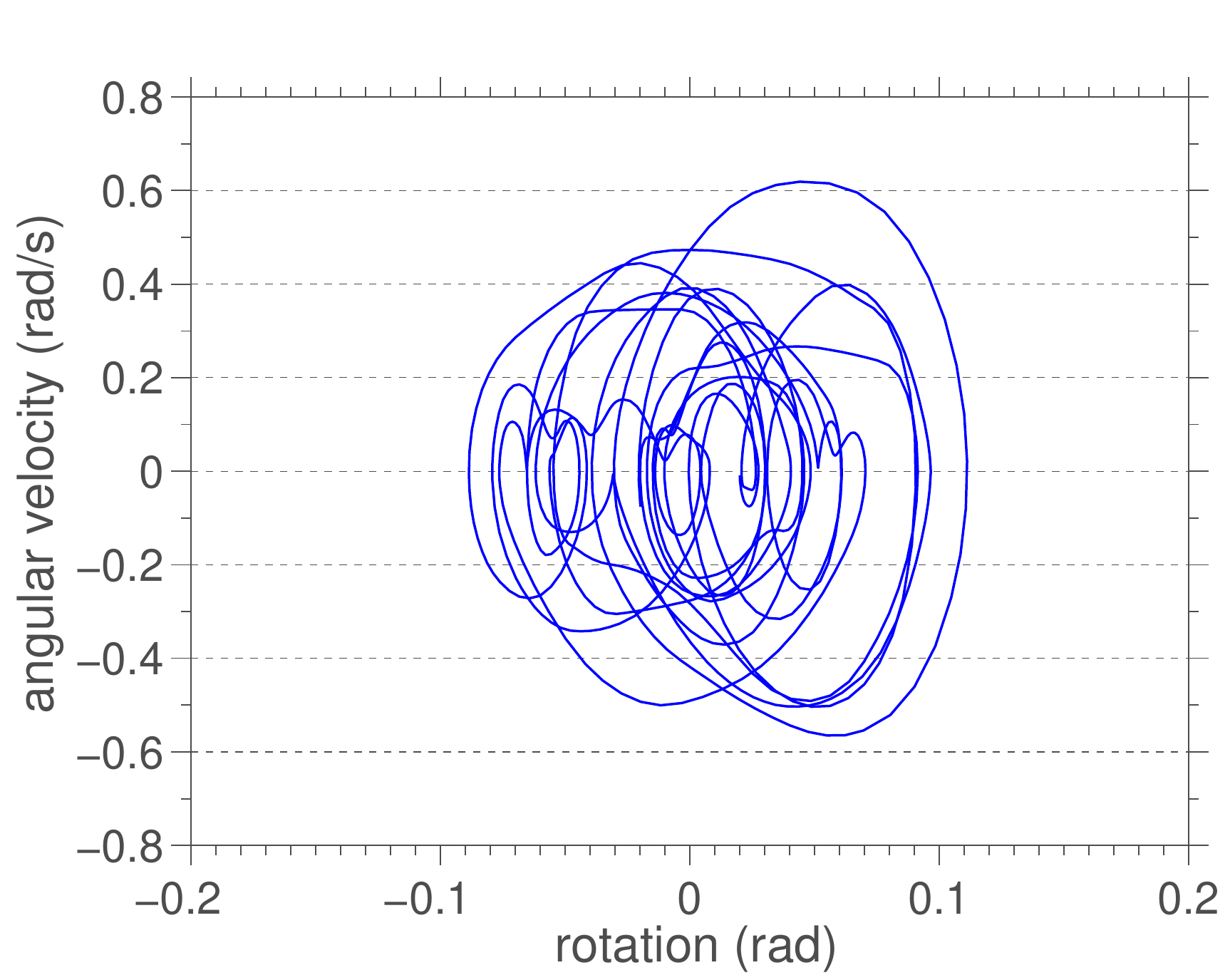}}
	\subfigure[]{	\includegraphics[scale=0.34]{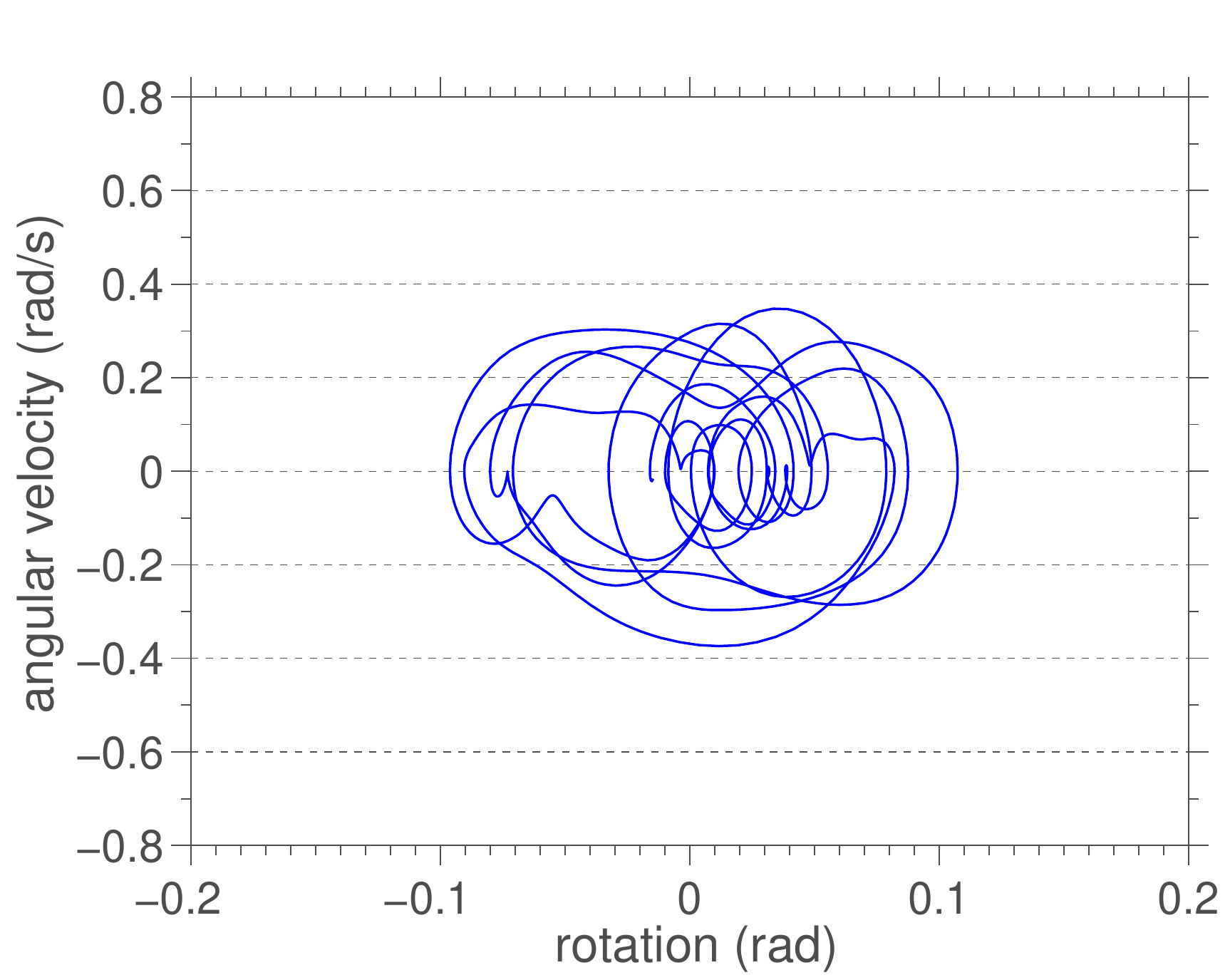}}
	\caption{Projections of rotational dynamics phase space trajectories.
	(a) $\phi_1$ atractor in $\R^3$; (b) $\phi_2$ atractor in $\R^3$; 
	(c) $\phi_1$ atractor in $\R^2$; (d) $\phi_2$ atractor in $\R^2$.}
	\label{phase_space_phi1_phi2_fig}
\end{figure}

The time series corresponding to the trailer/tower rotational dynamics 
$\phi_1$/$\phi_2$ is available in Figure~\ref{tseries_phi1_phi2_fig},
and the respective phase space trajectories projections are shown in 
Figure~\ref{phase_space_phi1_phi2_fig}.

Observe that in Figure~\ref{tseries_phi1_phi2_fig} the correlation between the 
time series is still strong, but there is a kind of filter effect, which
can also be noticed in the projected trajectories in Figure~\ref{phase_space_phi1_phi2_fig}.
Such trajectories seem to accumulate into a strange attractor, which justifies the 
irregular and intermittent appearance of the corresponding time series. However,
unlike for $y_1$ and $y_2$, now the difference between the two time series 
is not negligible. This significant difference between the rotations is responsible 
for nonlinear effects of inertia and damping. This is clear when looking at the 
matrices of Eqs.(\ref{matrix_M_eq}) and (\ref{matrix_N_eq}), which have trigonometric
terms that depend on $\phi_2 - \phi_1$.

\begin{figure}[h!]
	\centering
	\subfigure[]{\label{tseries_x2_figA} \includegraphics[scale=0.35]{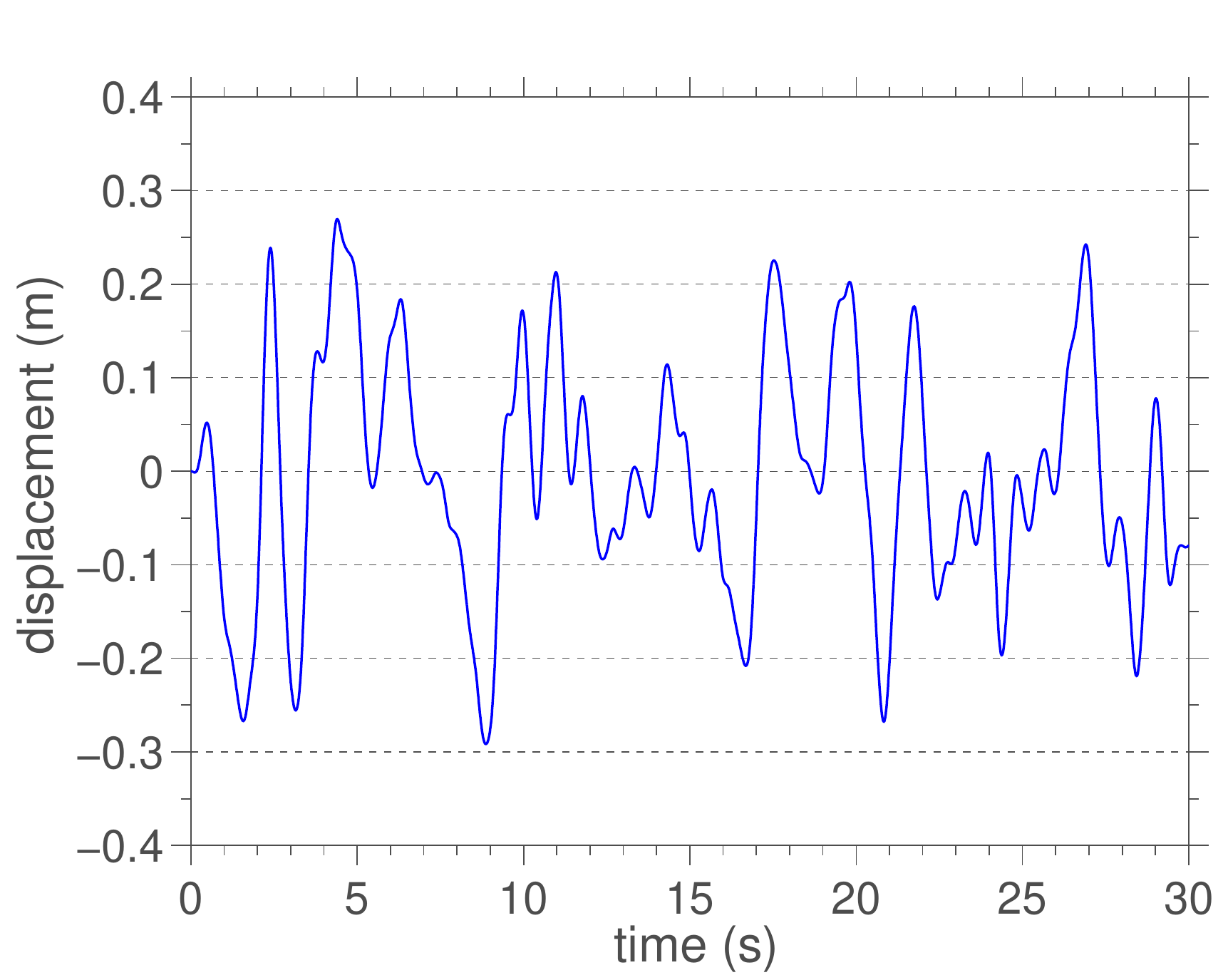}}\\
	\subfigure[]{\label{tseries_x2_figB} \includegraphics[scale=0.35]{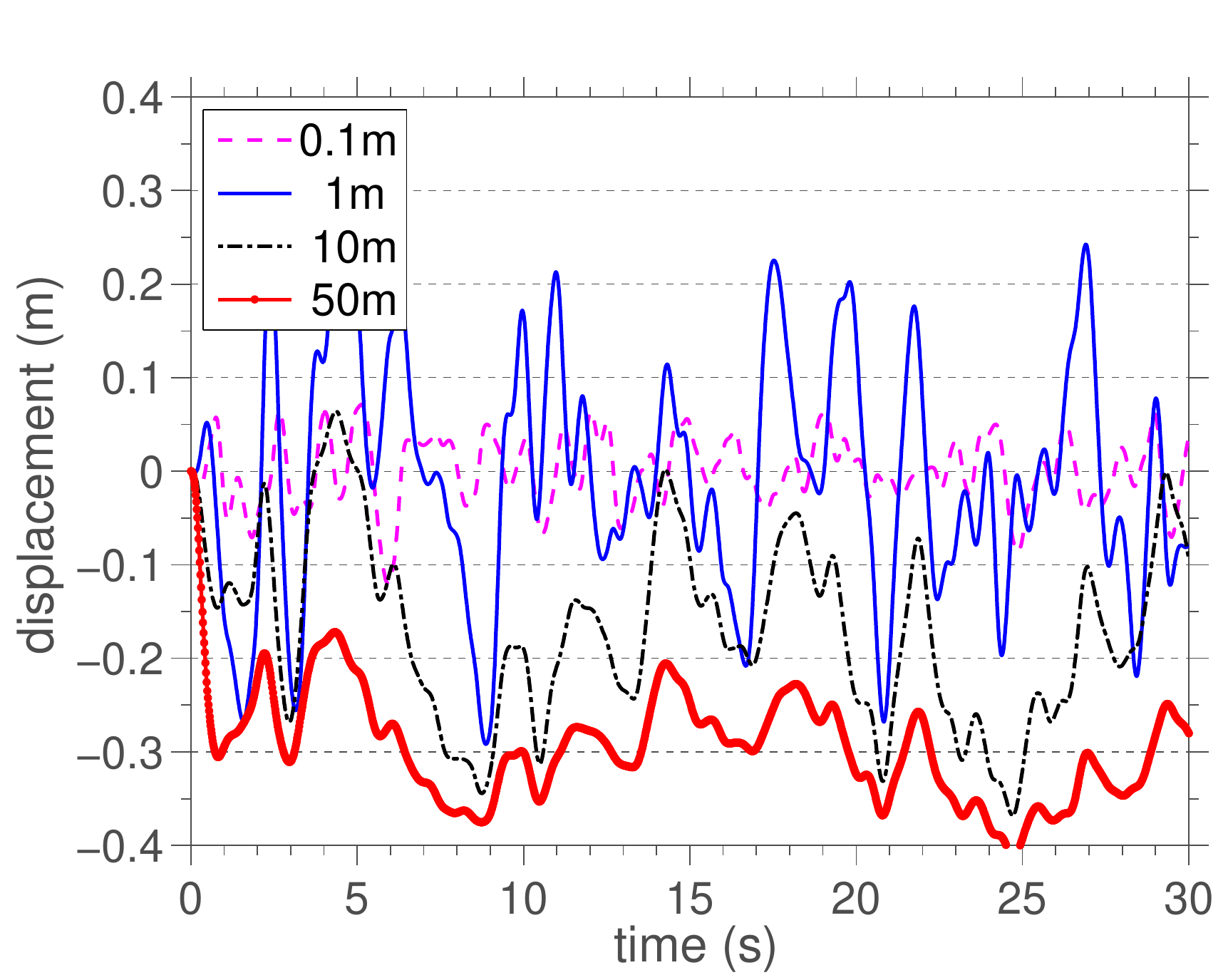}}
	\subfigure[]{\label{tseries_x2_figC} \includegraphics[scale=0.35]{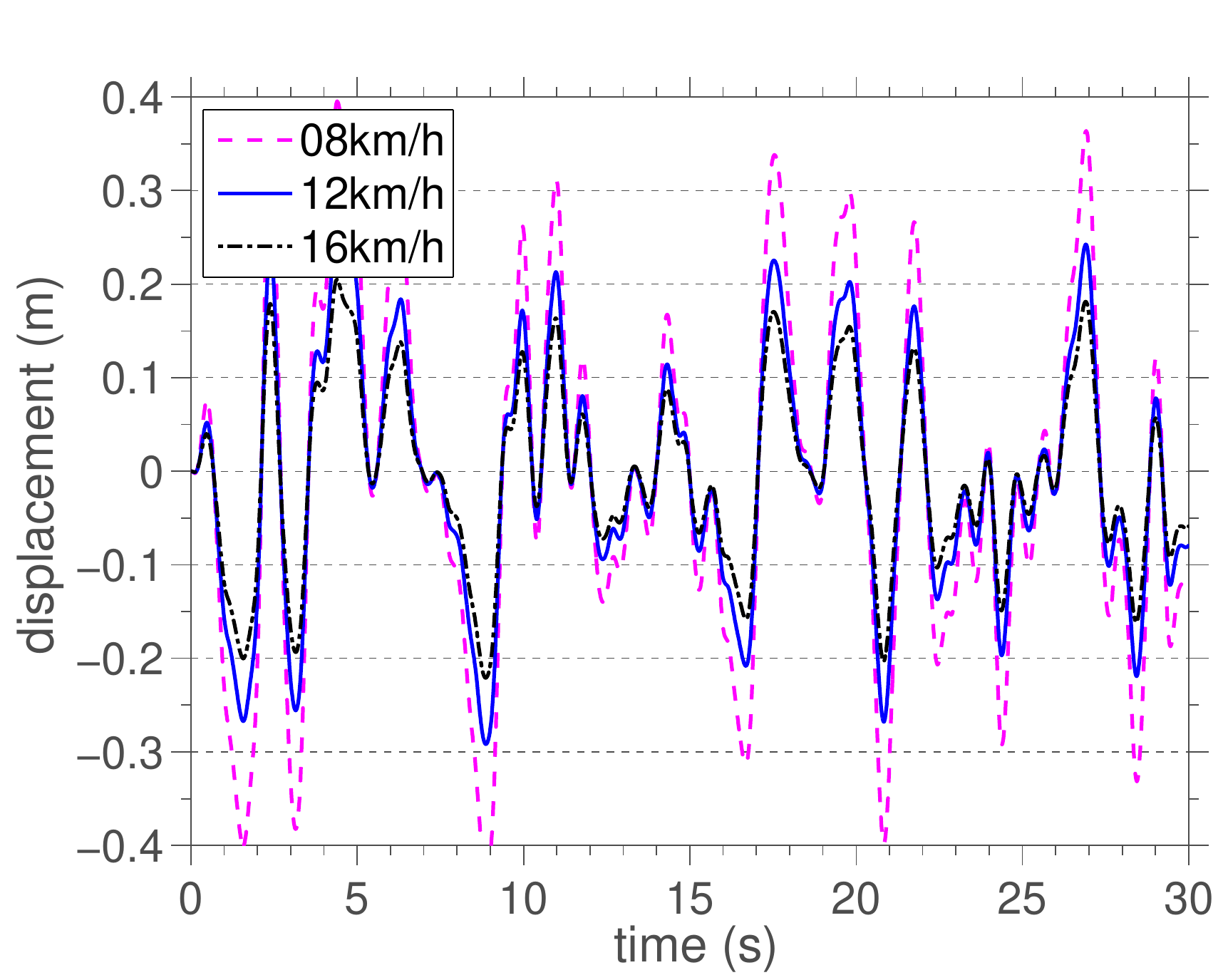}}
	\caption{Time series of tower horizontal dynamics $x_2$. (a) nominal set of parameters;
	(b) several values of $a_{corr}$; (c) several values of $v$.}
	\label{tseries_x2_fig}
\end{figure}

\begin{figure}[h!]
	\centering
	\subfigure[]{	\includegraphics[scale=0.35]{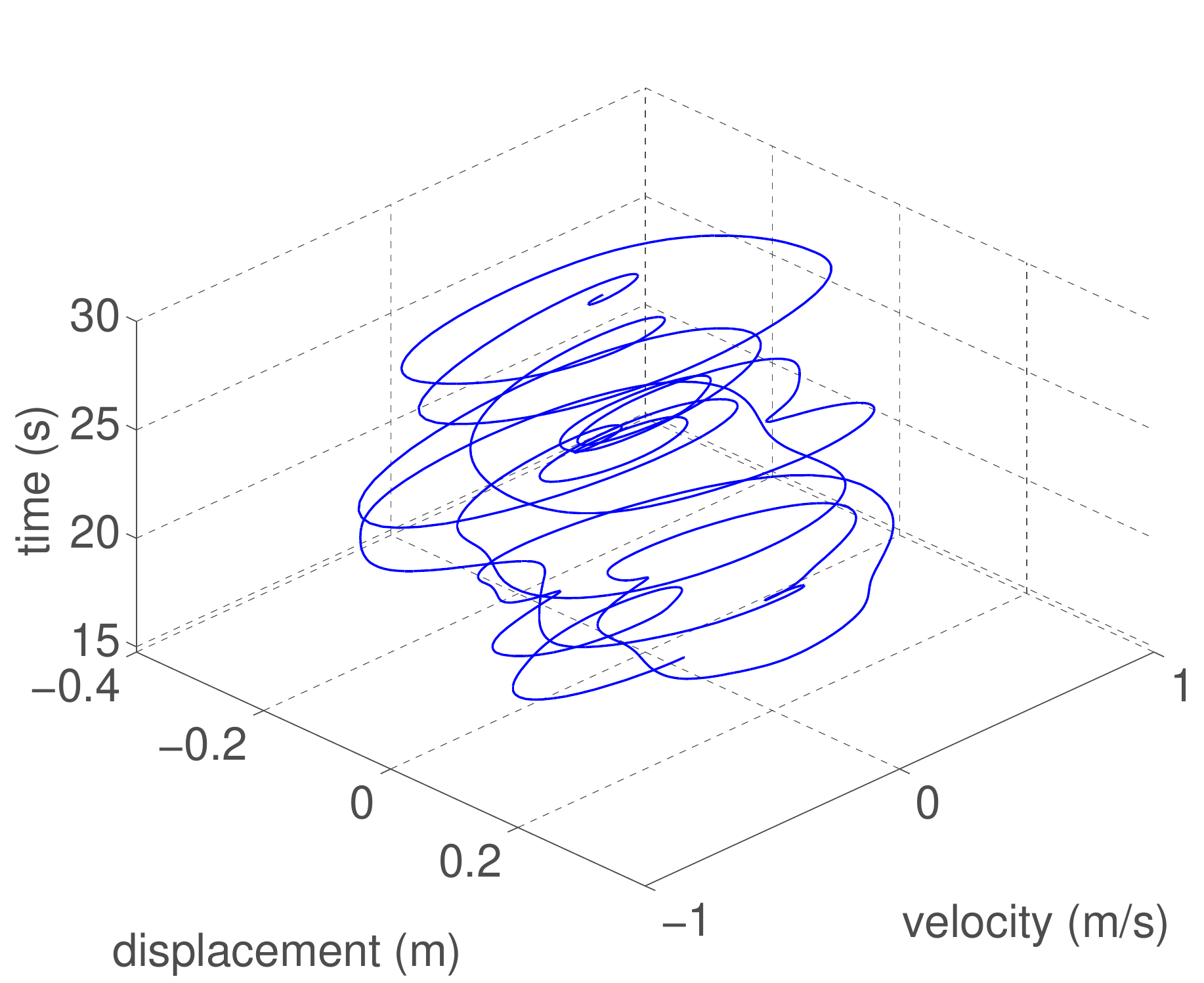}}
	\subfigure[]{	\includegraphics[scale=0.35]{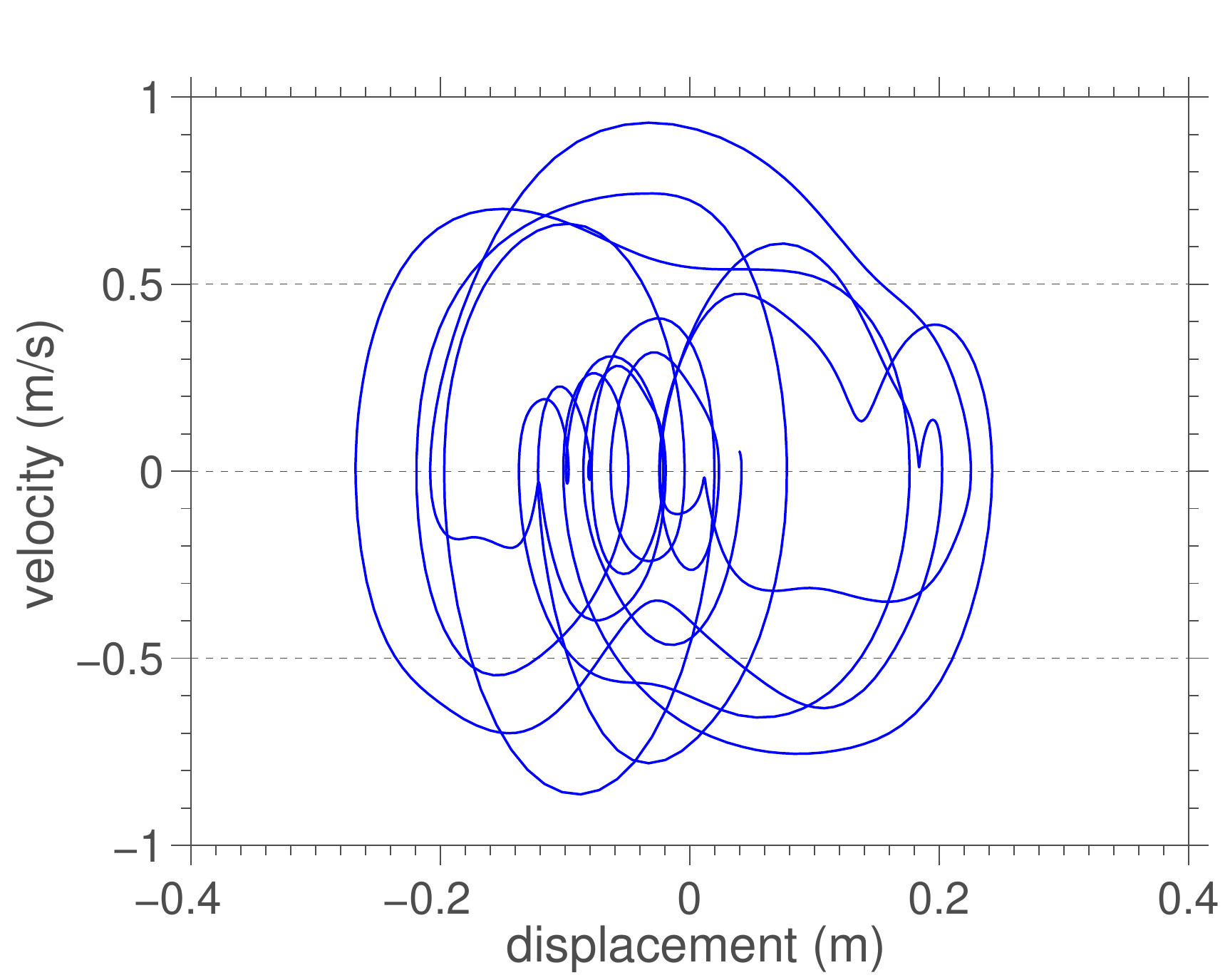}}
	\caption{Projections of horizontal dynamics phase space trajectory.
	(a) $x_2$ attractor in $\R^3$; (b) $x_2$ attractor in $\R^2$.}
	\label{phase_space_x2_fig}
\end{figure}

Finally, the time series corresponding to the tower horizontal dynamics $x_2$,
which is the main QoI associated to the dynamic system under study, is shown
in Figure~\ref{tseries_x2_figA}, and the associated phase space trajectories projections
are presented in Figure~\ref{phase_space_x2_fig}. An irregular dynamics
that accumulates into a strange attractor is noticed once more.

A parametric study on the behavior of $x_2$, for different values of correlation 
length $a_{corr}$ and translation velocity $v$ can be seen in 
Figures~\ref{tseries_x2_figB} and \ref{tseries_x2_figC}, respectively. Note that lateral 
oscillation amplitude strongly depends on the correlation length. This amplitude 
also depends on the translation velocity, in a way that it decreases as $v$ increases,
but this dependence is weaker than the one with $a_{corr}$.

\subsection{Spectral analysis}

In order to perform a spectral analysis of the dynamics, an estimation
of the power spectral density (PSD) of the QoI signal is constructed using 
the periodogram method \cite{oppenheim2009,soize1997}. In this algorithm, 
a wider interval of analysis is considered, for instance $[t_0,t_f]=[0,6000]$ s, 
so that QoI time series is segmented into non-overlying windows, being the 
signal PSD constructed through an averaging process, which uses estimations 
of the PSD for each window of the segmented signal.

\begin{figure}[h!]
	\centering
	\subfigure[]{\includegraphics[scale=0.35]{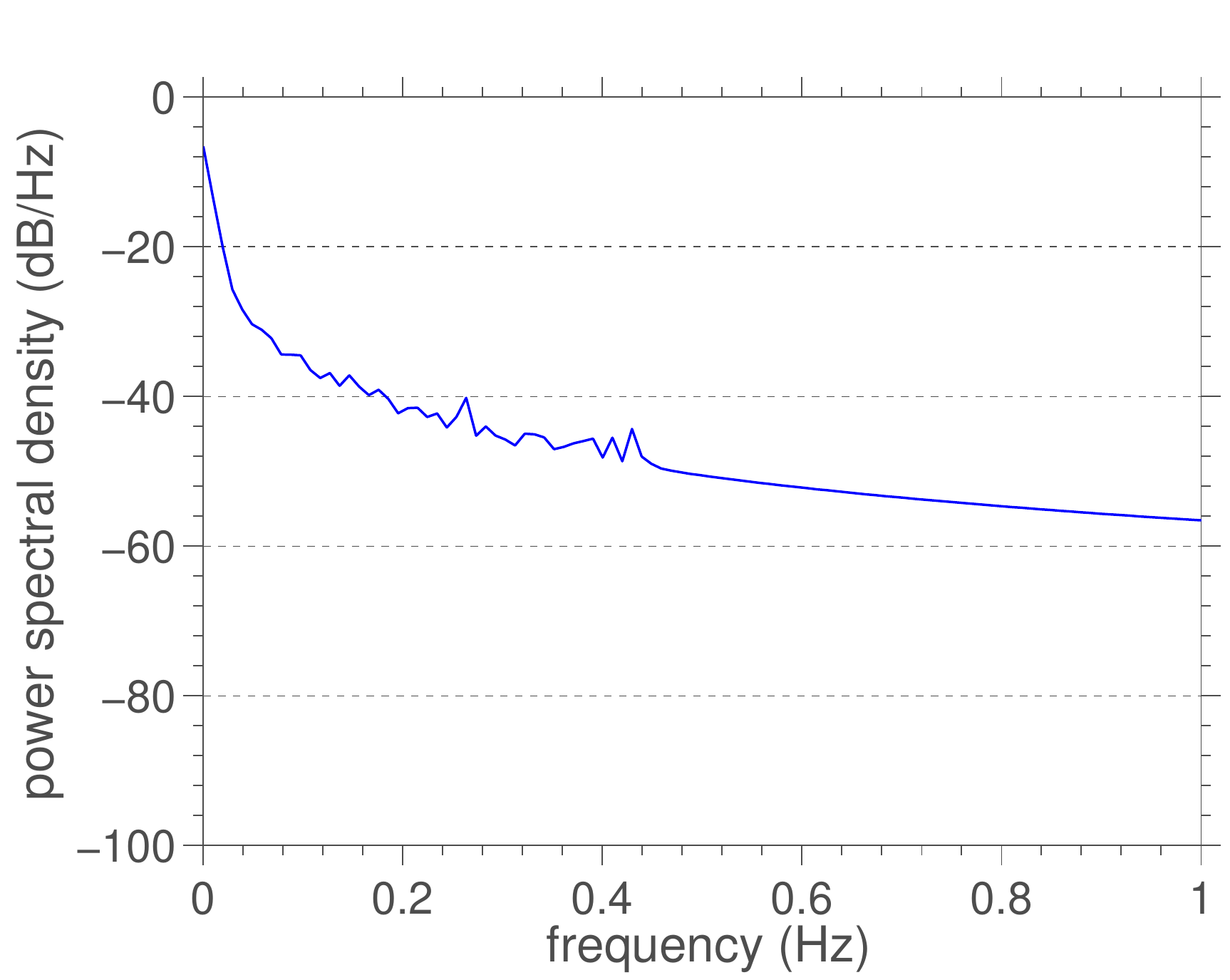}}
	\subfigure[]{\includegraphics[scale=0.35]{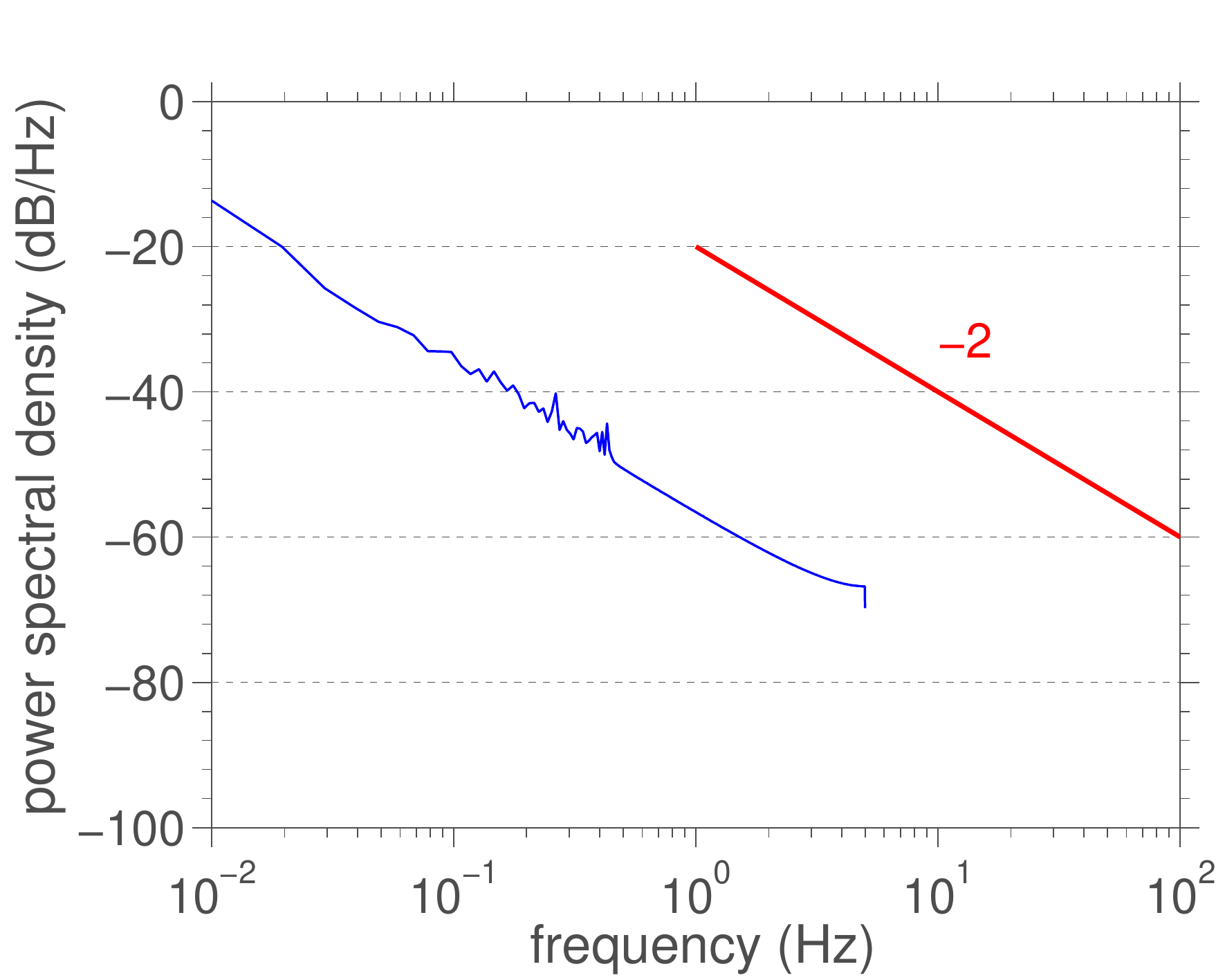}}
	\caption{Power spectral density of tower horizontal displacement. 
	(a) frequencies in linear scale. (b) frequencies in logarithmic scale.}
	\label{fourier_x2_disp_fig}
\end{figure}

The PSD of tower horizontal displacement signal is presented in
Figure~\ref{fourier_x2_disp_fig}, where it is possible to see that,
for almost all the frequencies in the band of interest, $\mathcal{B} = [0,5]$ Hz, 
the signal energy follows a linear decreasing law, with inclination -2.
This behavior, in form of a direct energy cascade, indicates that energy 
is injected in the system at the low frequencies of $\mathcal{B}$, 
being transferred in a nonlinear way through intermediate frequencies, 
until it is dissipated at the large frequencies by the structural damping.

\subsection{Convergence of MC simulation}

In order to ensure the ``quality" of the statistics obtained from
MC data, it is necessary to study the convergence of these 
stochastic simulations. For this purpose, it is taken into 
consideration the map $\texttt{conv}$, defined in 
section~\ref{MC_method}.

\begin{figure}
				\centering
				\includegraphics[scale=0.38]{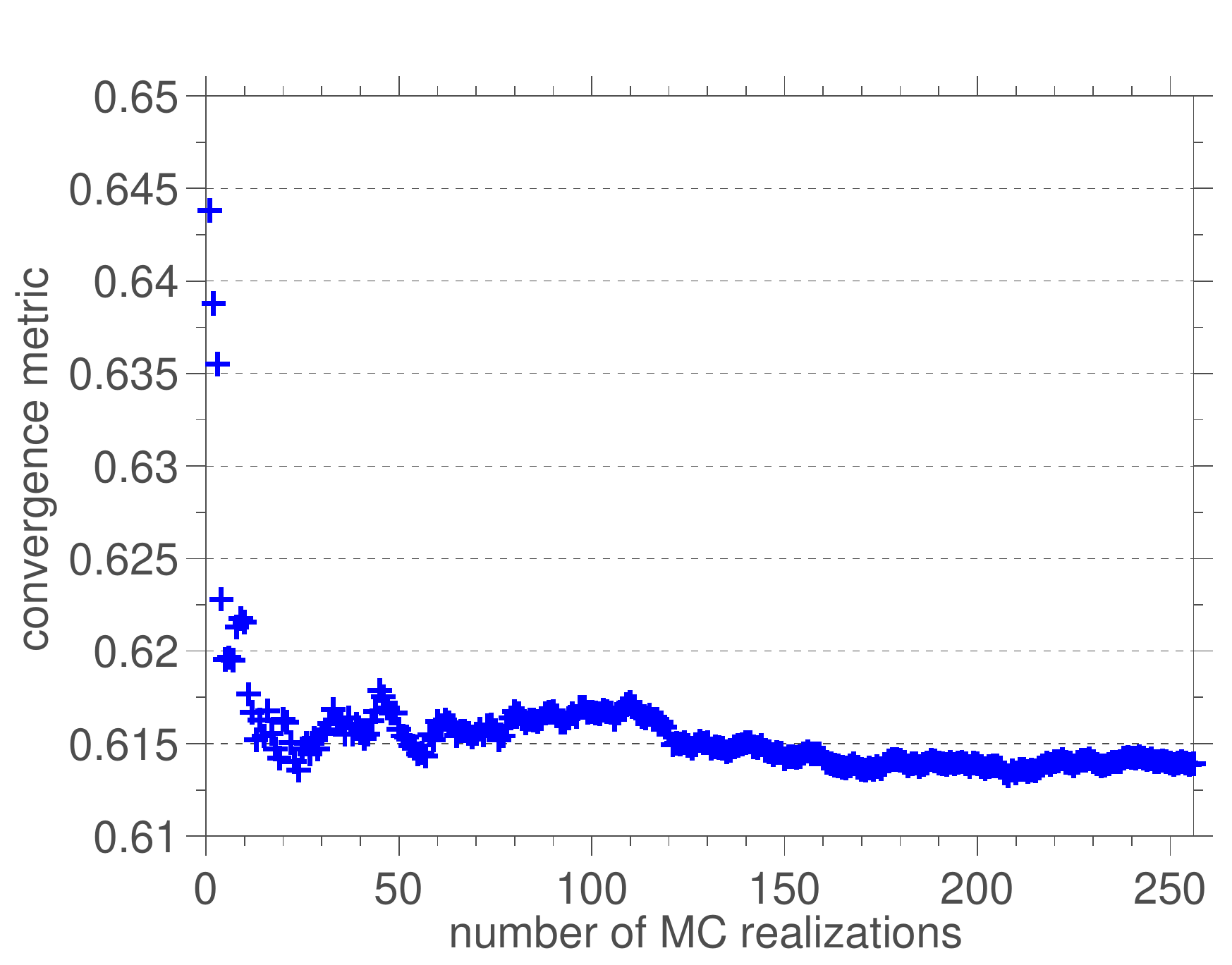}
				\caption{MC convergence metric as function of the realizations number.}
				\label{MC_conv_fig}
\end{figure}

The evolution of $\texttt{conv}(n_s)$ as a function of
$n_{s}$ can be seen in Figure~\ref{MC_conv_fig}. Note that for
$n_{s}=256$ the metric value has reached a steady value. 
So, all the stochastic simulations reported in this work use 
$n_{s}=256$.

Being the representative of the MC simulation guaranteed,
an analysis of how uncertainties (due to randomness in the 
external loading) are propagated through the model is
presented in the next sections.


\subsection{Confidence band and low order statistics}

\begin{figure}[h!]
	\centering
	\includegraphics[scale=0.38]{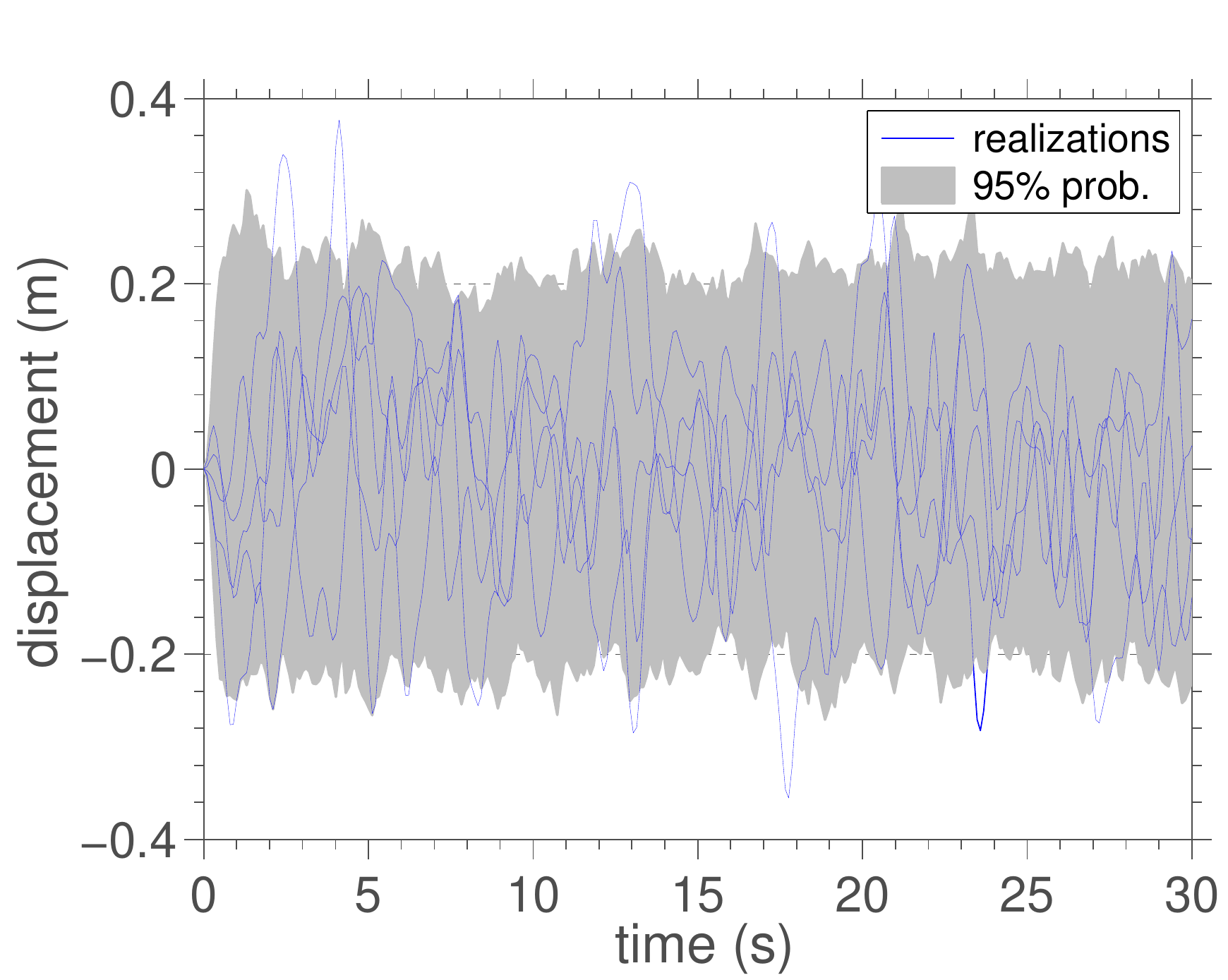}
	\caption{Confidence envelope and some realizations for tower horizontal displacement.}
	\label{cb_x2_disp_fig}
\end{figure}

\begin{figure}[h!]
	\centering
	\includegraphics[scale=0.38]{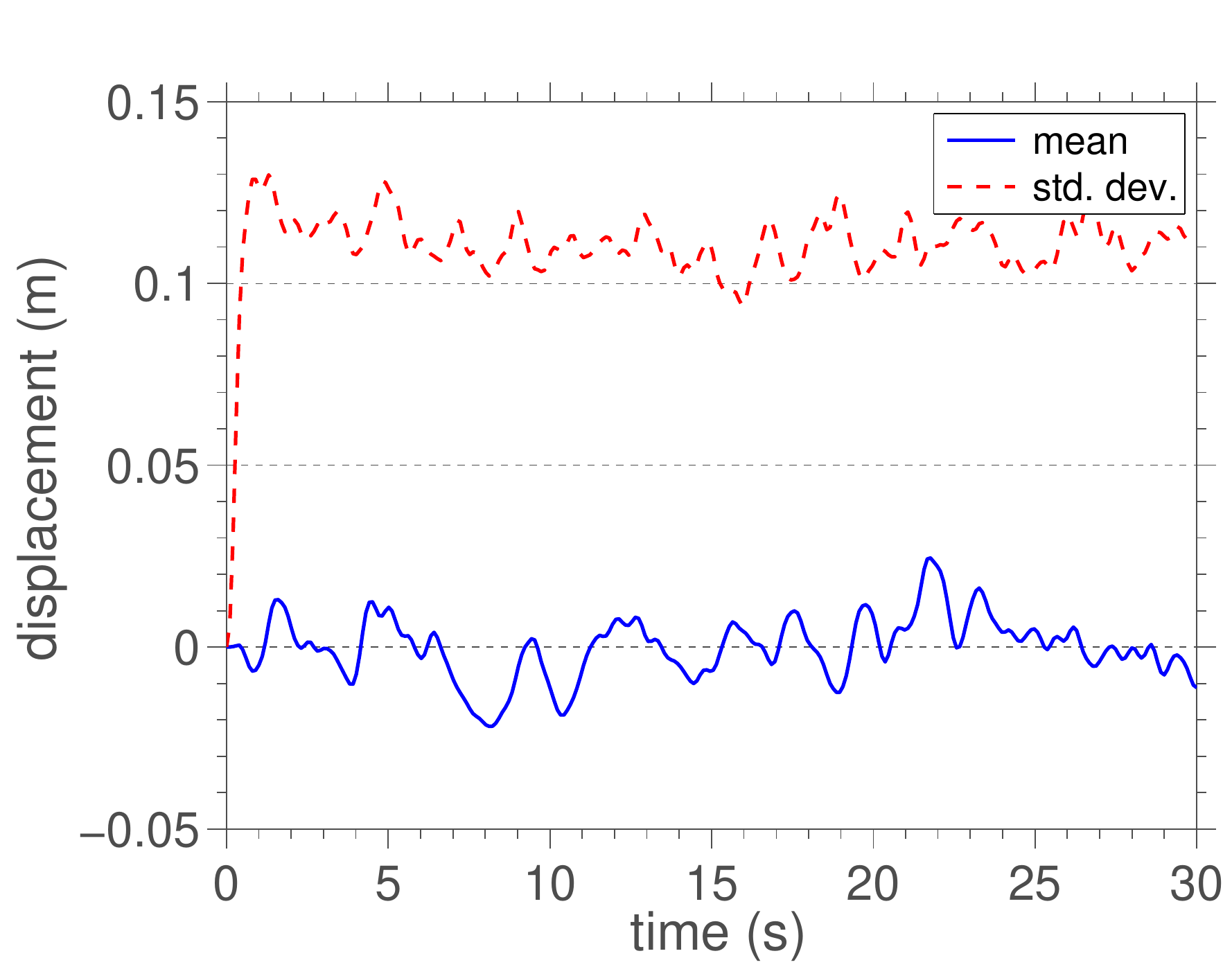}
	\caption{Sample mean/standard deviation for tower horizontal displacement.}
	\label{st_x2_disp_fig}
\end{figure}

In Figure~\ref{cb_x2_disp_fig} are presented some realizations of 
tower horizontal displacement and the corresponding confidence band
(grey shadow), wherein a realization of the stochastic system has 
95\% of probability of being contained. A wide variability in 
the QoI form can be observed. This fact may also be noted 
in Figure~\ref{st_x2_disp_fig}, which shows the evolution of the 
QoI sample mean and standard deviation. Note that $x_2$ has 
mean value near zero, but significant variability 
near all the interval of analysis.

\subsection{Evolution of tower horizontal vibration PDF}

\begin{figure}[h!]
	\centering
	\subfigure[$t=7.5\,s$]{\includegraphics[scale=0.35]{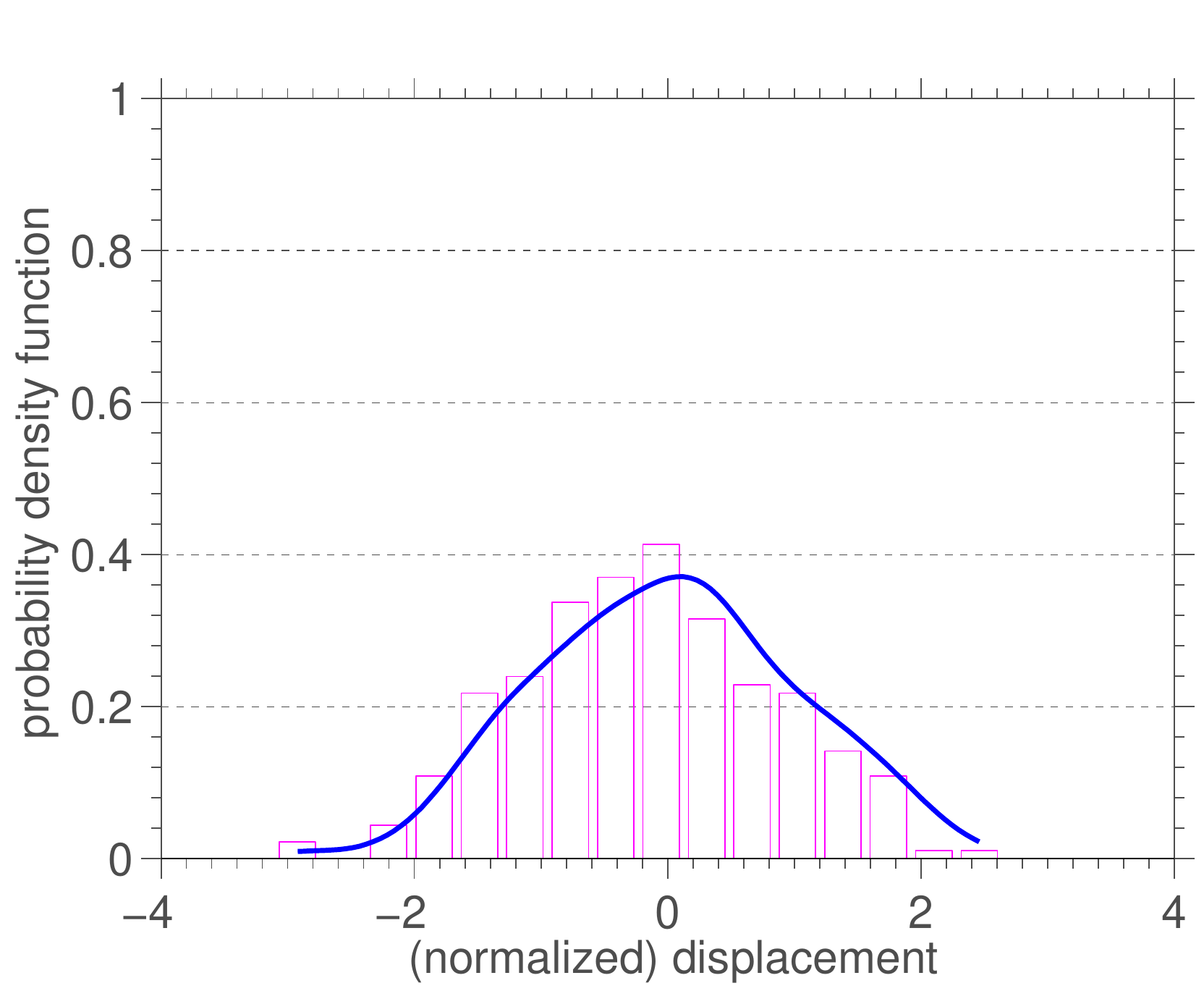}}
	\subfigure[$t=15.0\,s$]{\includegraphics[scale=0.35]{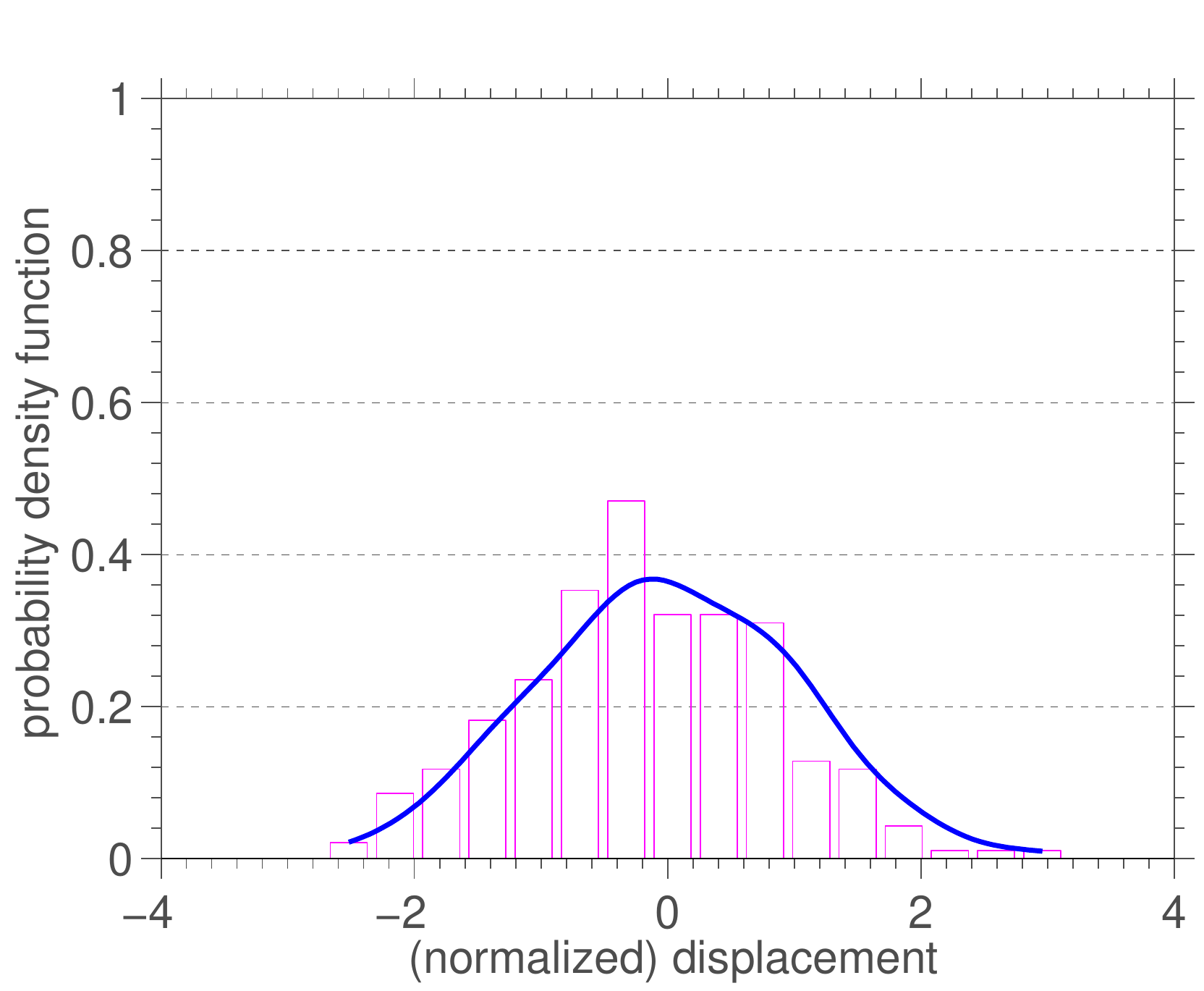}}\\
	\subfigure[$t=22.5\,s$]{\includegraphics[scale=0.35]{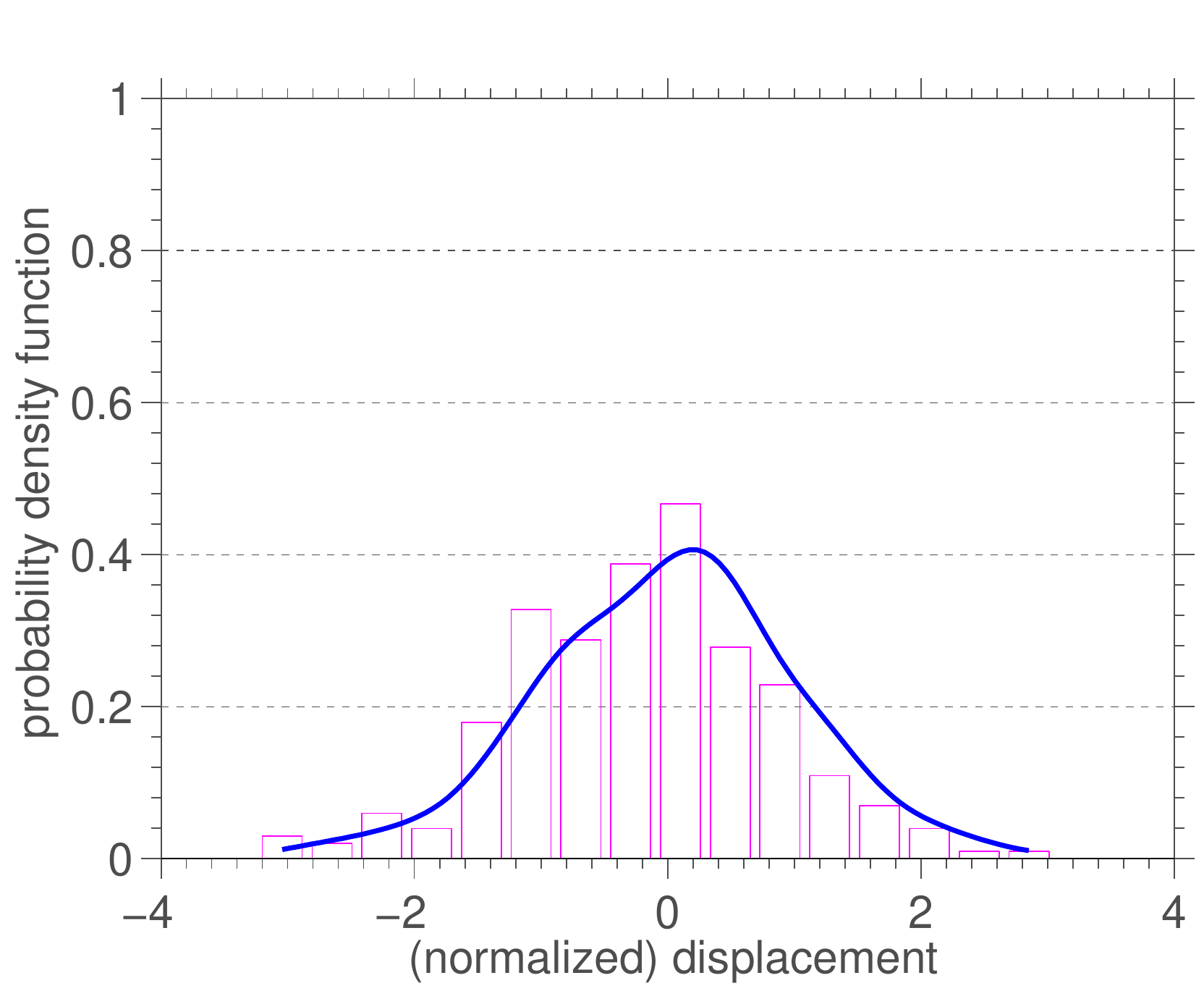}}
	\subfigure[$t=30.0\,s$]{\includegraphics[scale=0.35]{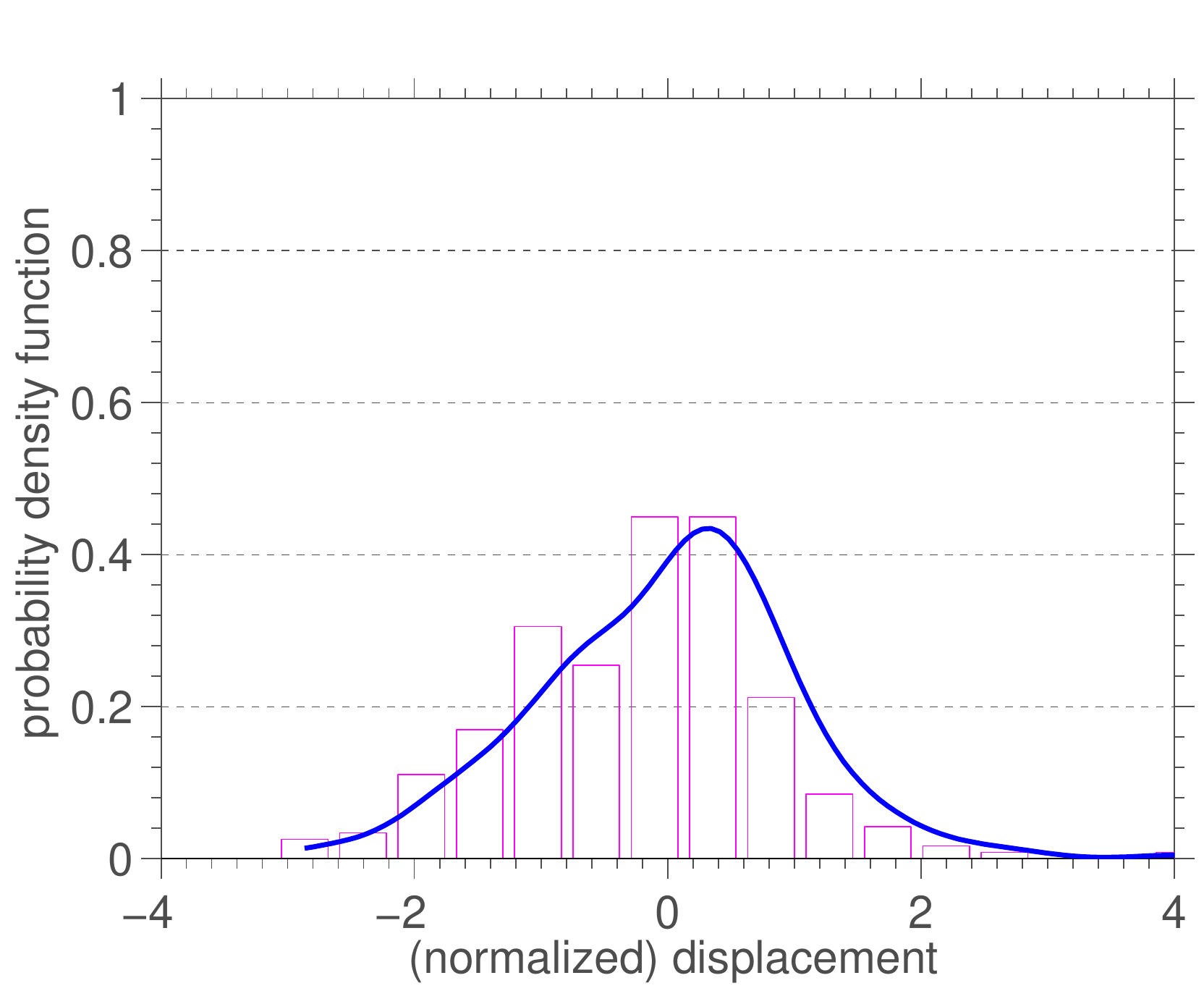}}
	\caption{Probability density function of tower horizontal dynamics (at different instants).}	
	\label{pdf_x2_fig}
\end{figure}

\begin{figure} [h!]
	\centering
	\includegraphics[scale=0.38]{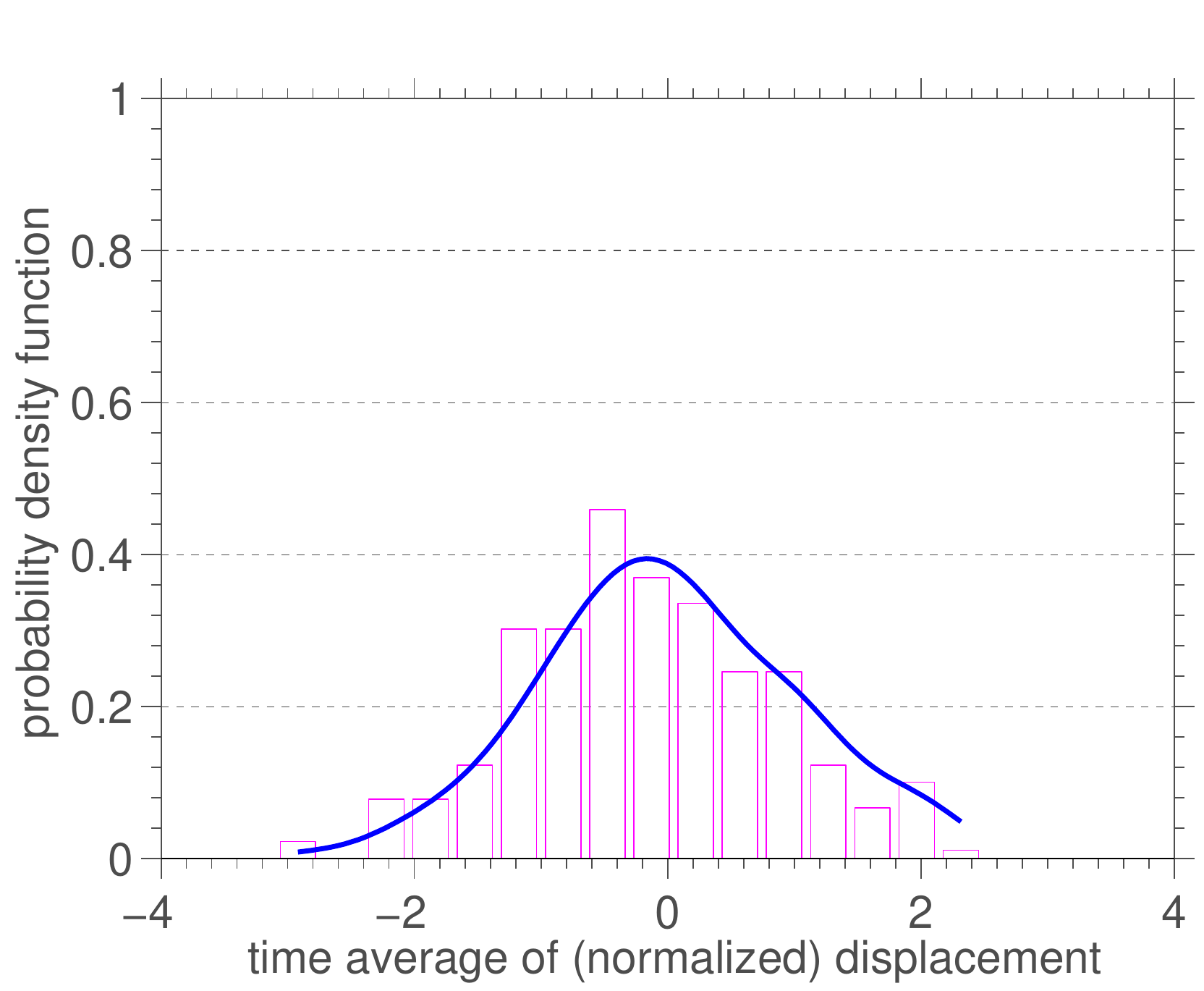}
	\caption{Time average of tower horizontal dynamics probability density function.}
	\label{mean_pdf_x2_fig}
\end{figure}

Estimations for the normalized\footnote{In this context, the meaning 
of normalized is zero mean and unity standard deviation.} probability 
density function (PDF) of the tower horizontal vibration, for different instants 
of time, are presented in Figure~\ref{pdf_x2_fig}.
In all cases it is possible to observe small asymmetries with respect to mean
and unimodal behavior, with maximum always occurring in the neighborhood
of the mean value. The time average of the tower horizontal dynamics PDF
is shown in Figure~\ref{mean_pdf_x2_fig}, which reflects the unimodal
characteristic of $x_2$ distribution observed in the time instants of
Figure~\ref{pdf_x2_fig}.

\subsection{Large vibrations probability}

When the uncertainties in orchard sprayer dynamics are quantified
through a probabilistic approach, it is of particular interest to calculate 
the probability of occurrence of extreme events. For instance, the structure 
presents large lateral (horizontal) vibrations due to some extreme loading 
or as a consequence of nonlinear interactions between the soil irregularities 
and the tower sprayer.

To perform this calculation it is necessary to define a level of lateral vibration 
that is considered high. In this paper, this value corresponds
to an amplitude of lateral vibration greater than 30\% of the distance between 
the left wheel and the trailer center line, i.e.,

\begin{equation}
	\text{large vibration} = \left\lbrace \mid\randvar{x}_2(t)\mid ~>~ \mbox{30\% of $B_1$} \right\rbrace.
\end{equation}

Therefore, for any instant $t$, it is of interest to determine the value of

\begin{equation}
	\PM \left\lbrace \mid \randvar{x}_2(t)\mid  ~>~ \mbox{30\% of $B_1$} \right\rbrace = 
	1 - \PM \left\lbrace \mid \randvar{x}_2(t)\mid  ~\leq~ \mbox{30\% of $B_1$} \right\rbrace,
	\label{prob_large_vib1}
\end{equation}

\noindent
where

\begin{equation}
	\PM \left\lbrace \mid \randvar{x}_2(t)\mid  ~\leq~ \mbox{30\% of $B_1$} \right\rbrace 
	= \int_{-0.3\,B_1}^{+0.3\,B_1} d \cdf{\randvar{x}_2(t)}{x_2}.
	\label{prob_small_vib1}
\end{equation}

\begin{figure} [h!]
	\centering
	\includegraphics[scale=0.38]{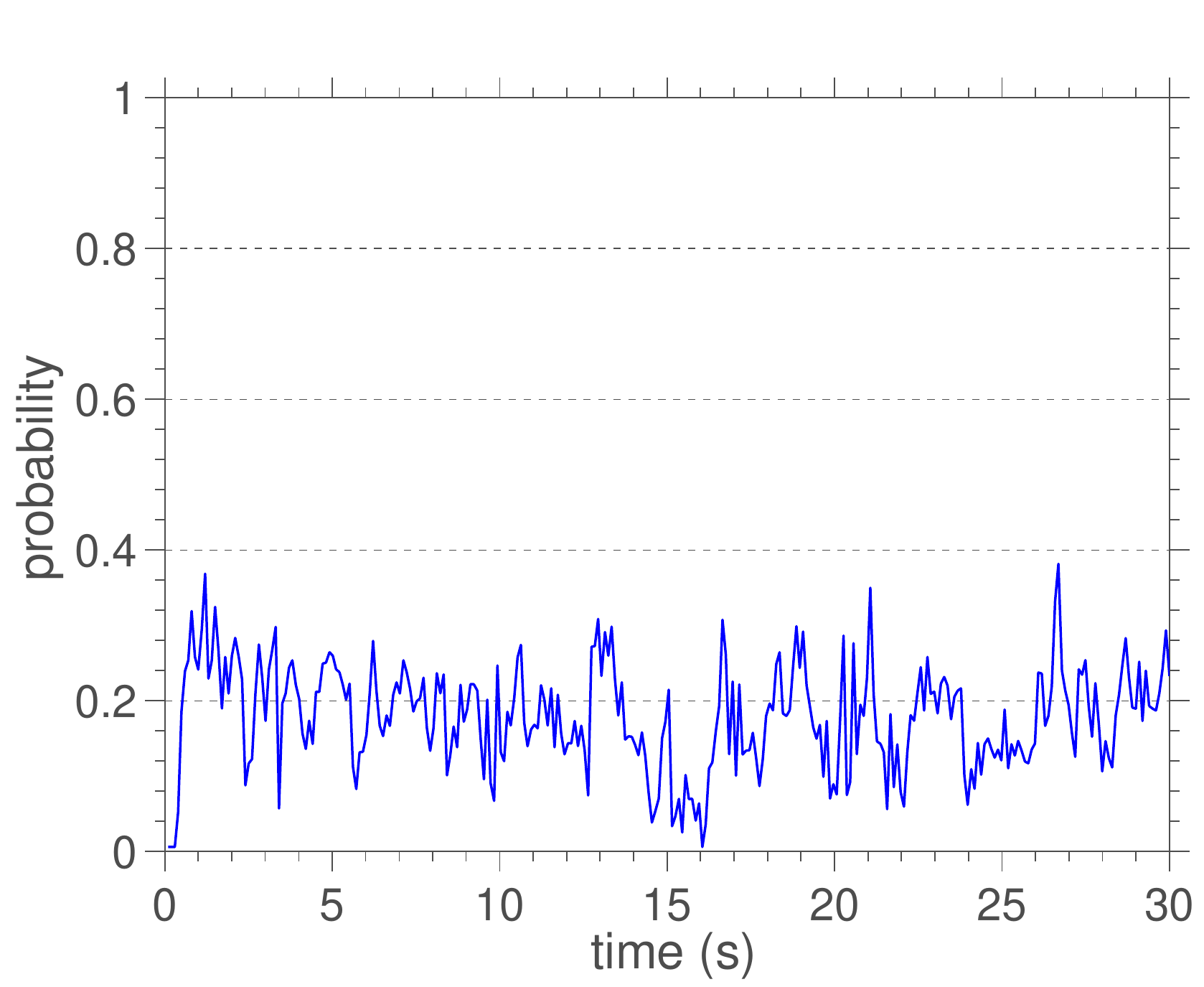}
	\caption{Evolution of the probability of large horizontal vibrations.}
	\label{prob_Qx2_fig}
\end{figure}

The last integral corresponds to the area of the curve that represents 
the PDF of $x_2(t)$, above the interval $\left[- 0.3\,B_1, 0.3\,B_1\right]$.
By calculating the integral of Eq.(\ref{prob_small_vib1}) for every instant in
the time interval of analysis, and then replacing the result in Eq.(\ref{prob_large_vib1}), 
the probability of large lateral vibrations as a function of time is obtained.
The evolution of this probability is shown in Figure~\ref{prob_Qx2_fig}, where 
the reader can note that the probability of an unwanted level of lateral 
vibration is not negligible in general (20\% on average), with peaks values
near 40\%.

\section{Conclusions}
\label{concl_remaks}

This work presented the study of the nonlinear dynamics of an orchard 
tower sprayer that is subjected to random excitations due to soil irregularities, 
modeled as a three degrees of freedom multibody system. The random loadings 
were taken into account through a parametric probabilistic approach, where the 
external loading was assumed to be a random process that is represented
through Karhunen-Lo\`{e}ve decomposition. The paper not only proposes 
a mechanical analysis, but also provides a formulation for this type of system, 
and a methodology, that can be reused to analyze other industrial equipment 
excited by soil irregularities induced loads.

Numerical simulations showed that orchard tower sprayer has a very rich nonlinear 
dynamics, which is able to reproduce complex phenomena such as chaos. 
The study also indicated that system dynamics follows a direct energy cascade law,
where the energy injected at the low frequencies are transferred into a nonlinear way 
through the middle frequencies of the band, being dissipated at the high frequencies.

A probabilistic analysis discloses a wide range of possible responses for the 
mechanical system, and shows a non negligible possibility of large lateral 
vibrations being developed during the sprayer operation. 

In a future work, the authors intend to address the problem of lateral vibrations 
from the robust optimization point of view. Using the stochastic model developed 
in this work, they intend to find a robust strategy to change pair of system parameters 
(e.g., torsional stiffness and damping) in a way the levels of lateral
vibrations are reduced. In parallel, it would be interesting to study 
the possibility of internal resonances, and to verify how they may disrupt or help 
in the operation of sprayer tower.


\section*{Author contributions}

J.M.B. proposed the problematic. A.C. and J.M.B. designed the research plan. 
J.M.B and J.L.P.F. constructed the deterministic model, based on a previous 
work of then. A.C. constructed an original stochastic model. J.L.P.F. implemented
a deterministic version of the computational code, which was extended and 
adapted to run stochastic simulations by A.C. Numerical simulations were carried 
out by A.C.. All the authors interpreted  and discussed the results. A.C. wrote the 
manuscript, which was then revised by J.L.P.F. and J.M.B.. All authors approved 
the final manuscript.

\section*{Acknowledgments}

The authors are indebted to the Brazilian agencies
CNPq (National Council for Scientific and Technological Development), 
CAPES (Coordination for the Improvement of Higher Education Personnel) 
and FAPERJ (Research Support Foundation of the State of Rio de Janeiro)
for the financial support given to this research. The help of Prof. Michel 
Tcheou (UERJ) with signal processing issues was of great value for this work.
They are also grateful to M\'{a}quinas Agr\'{\i}colas Jacto S/A, for the 
important data supplied, and to the anonymous referees’, 
for their useful comments and suggestions.

\section*{References}



\end{document}